\documentclass[%
 aip,
 amsmath,amssymb,
 reprint,%
]{revtex4-1}
\usepackage{amsmath}
\usepackage{amsthm}
\newtheorem{theorem}{Theorem}
\usepackage{graphicx}% Include figure files
\usepackage{dcolumn}% Align table columns on decimal point
\usepackage{bm}% bold math
\usepackage[utf8]{inputenc}
\usepackage[T1]{fontenc}
\usepackage{mathptmx}
\usepackage{etoolbox}
\usepackage{float}

\makeatletter
\def\@email#1#2{%
 \endgroup
 \patchcmd{\titleblock@produce}
  {\frontmatter@RRAPformat}
  {\frontmatter@RRAPformat{\produce@RRAP{*#1\href{mailto:#2}{#2}}}\frontmatter@RRAPformat}
  {}{}
}%
\makeatother
\begin{document}

\preprint{AIP/123-QED}

\title{H-Theorem and Boundary Conditions for Two-Temperature Model:\\Application to Wave Propagation and Heat Transfer in Polyatomic Gases}
% Force line breaks with \\
\author{Anil Kumar}
\email{p20200044@pilani.bits-pilani.ac.in}
 %\altaffiliation[Also at ]{}%Lines break automatically or can be forced with \\
\author{Anirudh Singh Rana}%
 \homepage{https://www.bits-pilani.ac.in/pilani/anirudhrana/profile}
 \email{anirudh.rana@pilani.bits-pilani.ac.in}
\affiliation{ 
Department of Mathematics \\Birla Institute of Technology and Science, Pilani, Rajasthan 333031, India%\\This line break forced with \textbackslash\textbackslash
}%

% \author{C. Author}
%  \homepage{https://www.bits-pilani.ac.in/pilani/anirudhrana/profile}
% \affiliation{%
% Second institution and/or address%\\This line break forced% with \\
% }%

\date{\today}% It is always \today, today,

\begin{abstract}
Polyatomic gases find numerous applications across various scientific and technological fields, necessitating a quantitative understanding of their behavior in non-equilibrium conditions. In this study, we investigate the behavior of rarefied polyatomic gases, particularly focusing on heat transfer and sound propagation phenomena. By utilizing a two-temperature model, we establish constitutive equations for internal and translational heat fluxes based on the second law of thermodynamics. A novel \textit{reduced two-temperature model} is proposed, which accurately describes the system's behavior while reducing computational complexity. Additionally, we develop phenomenological boundary conditions adhering to the second law, enabling the simulation of gas-surface interactions. 
The phenomenological coefficients in the constitutive equations and boundary conditions are determined by comparison with relevant literature. Our computational analysis includes conductive heat transfer between parallel plates, examination of sound wave behavior, and exploration of spontaneous Rayleigh-Brillouin scattering. The results provide valuable insights into the dynamics of polyatomic gases, contributing to various technological applications involving heat transfer and sound propagation.

\end{abstract}

\maketitle

% \begin{quotation}
% The ``lead paragraph'' is encapsulated with the \LaTeX\ 
% \verb+quotation+ environment and is formatted as a single paragraph before the first section heading. 
% (The \verb+quotation+ environment reverts to its usual meaning after the first sectioning command.) 
% Note that numbered references are allowed in the lead paragraph.
% %
% The lead paragraph will only be found in an article being prepared for the journal \textit{Chaos}.
% \end{quotation}

\section{\label{sec:level1}Introduction\protect\\}

Polyatomic gases have a plethora of applications that span across various fields of science and technology, ranging from aeronautics and astronautics to plasma physics and energy production. Understanding these phenomena in a quantitative and reliable manner is essential, as they present stimulating yet challenging scientific problems that garner significant research interest. 
Moreover, this understanding extends beyond gas flows, as demonstrated by the blue color of the sky and the red appearance of the moon during a lunar eclipse. These phenomena are attributed to Rayleigh scattering, occurring when particles in the atmosphere are smaller than the wavelength of light \cite{strutt1871xv}. Nitrogen and oxygen molecules, abundant in the Earth's atmosphere, play a major role in Rayleigh scattering. Investigating these scattering processes not only enhances our understanding of various atmospheric behaviors but also sheds light on optical phenomena influencing our environment.
\\
The intricate molecular structures found in polyatomic gases have a significant impact on the scattering of light and the resulting color observed. When designing gas sensors to detect and measure these types of gases effectively, it is essential to consider the strong non-equilibrium effects. This includes considering the limit of a large Knudsen number, $Kn$, which is determined by the ratio of the mean free path ($\lambda$) in the gas to a characteristic length scale ($L$) of the flow. Additionally, it is important to consider a large Weissenberg number \cite{kara2017generalized}, which relates to the relaxation time that characterizes the rate at which perturbations in the gas decay in relation to the characteristic time scale of the system, such as the inverse frequency of light or the sound wave. In particular, in polyatomic gases, another mechanism responsible for deviation from equilibrium is the finite rate of relaxation of internal degrees of freedom with random translational energy, which leads to a large ratio of bulk viscosity to shear viscosity. While the classical Navier-Stokes-Fourier (NSF) equations are applicable
when the ratio of bulk viscosity to shear viscosity is small, gases such as
$CH_4$, $CO_2$, and $H_2$ exhibit significantly large ratios \cite{sharma2022estimation,sharma2019estimation}, rendering the
one-temperature model is inadequate\cite{Kustova2023}. In such cases, the two-temperature model
becomes essential for accurately capturing the system's behavior \cite{aoki2021note,ruggeri2015rational,kustova2020multi}. 
\\
In recent literature, several sets of equations for the two-temperature model have been proposed and studied. Notably, the two-temperature Navier-Stokes equations derived from an ellipsoidal Bhatnagar-Gross-Krook (ES-BGK) model for a polyatomic gas have been studied \cite{aoki2020two}. These investigations focus on regimes where the bulk viscosity significantly exceeds the shear viscosity and are based on a discrete structure of internal energy levels \cite{bisi2019two,kustova2023continuum}. In the current study, we concentrate on determining the non-equilibrium distribution function using the maximum entropy principle\cite{dreyer1987maximisation} and establishing the second law of thermodynamics for the two-temperature model. The phenomenological coefficients are determined by
comparing different collision models found in the literature.
\\
In the field of nonequilibrium thermodynamics, various approaches exist to
determine the behavior of a system near equilibrium. One approach is
Linear Irreversible Thermodynamics (LIT) \cite{de2013non}, which assumes local thermodynamic
equilibrium and derives constitutive laws for the stress tensor and heat
flux based on the second law of thermodynamics.  On the other hand, Rational Extended Thermodynamics (RET) relaxes the requirement of local thermodynamic equilibrium by introducing the Clausius-Duhem equation as a specific form of entropy balance law \cite{noll1974thermodynamics}. 
% In RET, the non-convective entropy flux is defined as the heat flux divided by the thermodynamic temperature derived from LIT. By carefully analyzing conservation laws and the entropy equation, constitutive equations for the stress tensor and heat flux can be obtained. 
Despite their differing postulates, RET \cite{ruggeri2015rational, muller2013rational}  and LIT, two thermodynamic approaches, yield equivalent constitutive equations for simple fluids, including those governing local thermodynamic equilibrium \cite{ noll1974thermodynamics, muller1985thermodynamics}. An important feature of both approaches is that the entropy generation rate is expressed as the sum of products of thermodynamic forces and thermodynamic fluxes.  
The formulation begins by establishing an extended Gibbs equation and
applying the second law of thermodynamics to ensure a positive entropy
generation rate.  
\\
While thermodynamics, including both rational and extended versions, is commonly employed to investigate these phenomena, a kinetic approach \cite{chapman1990mathematical,cercignani2000rarefied} provides deeper insights, although it comes with the drawback of large computational costs. As a result, in many cases, it becomes desirable to utilize simpler macroscopical models, known as extended hydrodynamics models, which prove highly valuable for engineering purposes. Rahimi and Struchtrup \cite{rahimi2016macroscopic} have developed a kinetic model and a high-order macroscopic model to accurately represent rarefied polyatomic gas flows at moderate Knudsen numbers. The kinetic model extends the Shakov model (S-model) \cite{rykov1975model,shakhov1968generalization} and accurately captures the dynamics of higher moments. Through the order of magnitude method \cite{struchtrup2013regularized,struchtrup2004stable}, optimized moment definitions and scaled Grad's 36-moment equations are obtained. The first order yields a modified version of the Navier-Stokes-Fourier equations, while the third-order results in a set of 19 regularized partial differential equations (R19). Aoki et al. \cite{aoki2020two} investigated a polyatomic gas with slow relaxation of internal modes and derived the Navier-Stokes equations with two temperatures (translational and internal temperatures) based on the ellipsoidal-statistical (ES) model of the Boltzmann equation proposed by Andries et al. \cite{andries2000gaussian}. The derivation was carried out using the Chapman-Enskog procedure. Djordji\'{c} et al. \cite{djordjic2023boltzmann} developed collision kernel models and used the nonlinear Boltzmann collision operator for polyatomic gases to derive explicit expressions for transport coefficients, including shear and bulk viscosities and thermal conductivity. These coefficients depend on the parameters of the collision kernel. Marques and Kremer \cite{marques1993spectral} proposed a hydrodynamical model that incorporates a relaxation equation for dynamic pressure into the conventional hydrodynamic equations based on the field equations of a polyatomic gas consisting of rough spherical molecules \cite{kremer1987kinetic}.
\\
In this article, we first establish the second law for the two-temperature model by deriving the extended Gibbs equation from the maximum entropy distribution, incorporating six field variables, namely, density, momentum, translational temperature, and internal temperature. Furthermore, we derive constitutive equations for internal heat flux ($q_{k}^{tr}$) and translational heat flux ($q_{k}^{in}$) based on the second law. Additionally, drawing inspiration from the order of magnitude technique \cite{struchtrup2013regularized, struchtrup2004stable}, we reformulate the internal and translational heat fluxes as the summation of the total heat flux ($q_{k}$) and additional heat flux ($Q_{k}$), with the magnitude of $Q_{k}$ being larger than that of the total heat flux \cite{rahimi2016macroscopic}. Thus, to achieve the desired accuracy for the first order, we determine the phenomenological coefficients such that the impact of the additional heat flux vanishes. This approach is referred to as the ``reduced two-temperature model,'' where the relevant field variables include mass momentum and two temperatures, while the constitutive relationships are established for stress and the total heat flux. 
In order to assess the validity and scope of the two-temperature model  equation, we examine its performance in analyzing time-dependent phenomena such as sound propagation and light scattering in dilute polyatomic gases. By comparing our theoretical predictions to experimental data, including acoustic measurements in nitrogen and oxygen by Greenspan \cite{greespan1959rotational}, and the extended hydrodynamic theory of Hammond and Wiggins \cite{marques1999light} in methane, we demonstrate that the proposed model equation effectively describes the acoustic properties and light scattering spectrum of dilute polyatomic gases.
\\
Boundary conditions play a crucial role in gas dynamics simulations as they define the behavior of the fluid at the boundaries, describing how the gas interacts with solid surfaces. This includes various phenomena, such as momentum and heat transfer, chemical reactions, and phase change. The selection of suitable boundary conditions significantly impacts the accuracy and reliability of simulation outcomes. In this study, our objective is to derive phenomenological boundary
conditions (PBCs) specifically for the two-temperature model applied to polyatomic gases.
\\
To establish these PBCs, we employ an entropy balance integrated around the
interface between the solid and gas. The PBCs are developed as
empirical rules to ensure a positive entropy inequality at the boundary and
to represent the entropy generation at the boundary using these relations.
These PBCs can be employed to solve boundary value problems. In a recent
work by \cite{rana2016thermodynamically}, the authors proposed phenomenological boundary conditions for
the linearized R13 equation using the second law of thermodynamics. They
evaluated the phenomenological coefficients by comparing slip/jump and
thermal creep coefficients with the linearized Boltzmann equation for
different accommodation coefficients. Similarly, in a set of
phenomenological boundary conditions was proposed for a coupled constitutive
relation \cite{waldmann1967non, rana2018coupled}. These conditions were designed to uphold the second law of
thermodynamics, a fundamental principle in physics governing the behavior of
energy and heat. 
% Waldmann \cite{waldmann1967non} deserves credit for developing a
% comprehensive thermodynamic approach to phenomenological boundary
% conditions.
\\
Kosuge et al. \cite{kosuge2021boundary} developed slip boundary conditions for the two-temperature system with a polyatomic gas, utilizing the ES model and incorporating the Maxwell-type diffuse-specular reflection condition on the boundary. Rahimi \&
Struchtrup \cite{rahimi2016macroscopic} introduced a kinetic boundary condition that incorporates the
concept of two distinct exchanging processes: translational and internal.
They utilize this condition to derive appropriate macroscopic boundary
conditions.
\\
For both the two-temperature model and the reduced two-temperature model,
this article establishes a set of wall boundary conditions adhering to the
second law of thermodynamics. The phenomenological coefficients appearing in
the boundary conditions are calculated by comparing them with kinetic theory
in the asymptotic limit of small dynamic temperature ($\vartheta $) \cite{rahimi2016macroscopic}.
\\
This study analyzes conductive heat transfer in rarefied polyatomic gases confined between parallel plates.  Furthermore, we examine the behavior of sound waves in rarefied polyatomic gases, with a specific focus on nitrogen and hydrogen gases. Additionally, we explore the occurrence of spontaneous Rayleigh-Brillouin scattering. The exact investigation of the Rayleigh-Brillouin spectral line shape is of practical importance as it provides valuable information about the velocity, density, and temperature of gas samples when illuminated. Heat transfer configurations play a critical role in various technological applications, including vacuum pressure gauges \cite{jitschin2004dynamical}, vacuum solar collectors \cite{o1992experimental}, multilayer insulation blankets used in space and cryogenic equipment \cite{sun2009experimental}, as well as micro heat exchangers and microsensors \cite{yang2014design,chalabi2012experimental}. Furthermore, these configurations serve as standard setups for determining important properties such as the thermal conductivity of gases \cite{saxena1971transport}, temperature jump coefficient \cite{sharipov2005velocity}, and energy accommodation at different surface temperatures \cite{semyonov1984investigation}. These evaluations involve a combination of modeling and experimental measurements \cite{trott2011experimental,yamaguchi2012investigation}.
\\
The remaining sections of the paper are organized as follows: In \S\,\ref{Model equation}, we establish the definitions of moments and derive the conservation laws and extended balance equations from the Boltzmann equation. In \S\,\ref{The second law of thermodynamics}, we focus on determining the non-equilibrium distribution function by applying the maximum entropy principle and proving the second law of thermodynamics. Additionally, we derive constitutive relations for the two-temperature model and determine the values of phenomenological coefficients. Subsequently, we discuss linearized and dimensionless equations and introduce a reduced model in  \S\,\ref{Linearized and dimensionless equations} and \S\,\ref{Reduced model}, respectively. In \S\,\ref{Linear stability analysis}, we conduct a linear stability analysis for different values of phenomenological coefficients. Furthermore, we analyze sound wave propagation in \S\,\ref{Sound wave propagation} and explore the problems of spontaneous Rayleigh-Brillouin scattering in \S\,\ref{Spontaneous Rayleigh-Brillouin Scattering}. Determining  the appropriate wall boundary conditions for the two-temperature model using the second law of thermodynamics is addressed \S\,\ref{Wall Boundary Conditions}.  We validate the above wall boundary conditions through an investigation of the fundamental problem of heat transfer between two parallel plates in \S\,\ref{Flow between two parallel plates}. Finally, in \S\,\ref{Conclusions}, we present our concluding remarks.

\section{\label{Model equation}Moment system}
The Boltzmann equation provides a kinetic description of polyatomic gases via
the one-body distribution function $f(t,\mathbf{x},\mathbf{c},I)$ which can be
formally written as \cite{ruggeri2015rational}%
\begin{equation}
\frac{\partial f}{\partial t}+c_{k}\frac{\partial f}{\partial x_{k}}+F_{k}%
\frac{\partial f}{\partial c_{k}}=\mathcal{S}[f,f]\text{.} \label{Boltzman equation}
\end{equation}%
The distribution function $f(t,\mathbf{x},\mathbf{c},I)$ describes
the state of the gas molecules having three translational degrees of freedom
and internal degrees of freedom, where $t\in 
\mathbb{R}^{+}$ is the time, $\mathbf{x}\in 
\mathbb{R}^{3}$ is the spatial position, $\mathbf{c}\in 
\mathbb{R}^{3}$ is the molecular translational velocity, $F_k$ is the field of the external forces, e.g. gravity, and $I$ denotes the specific
the energy of a gas molecule due to internal modes with $I\in 
%TCIMACRO{\U{211d} }%
%BeginExpansion
\mathbb{R}
%EndExpansion
^{+}$. The right-hand side of (\ref{Boltzman equation}) is known as the
collision operator, which represents the rate of change in $f$ due to binary
collisions. The collision operator in (\ref{Boltzman equation}) involves
complex integrals whose actual form depends on the detailed nature of the
intermolecular interactions. Furthermore, the binary collision operator $\mathcal{S}
$ have five collision invariants: mass, three components of momentum, and energy, given by%
\begin{equation}
\psi =m\left\{ 1\text{, }c_{i}\text{, }\frac{C^{2}}{2}+I\right\},
\label{collision invariants}
\end{equation}%
i.e., $\int \mathcal{S}\psi d\mathbf{c}dI=0$.
\\
The macroscopic variables, namely local density $\rho $, momentum density $%
\rho v_{i}$, and internal energy $u$, can be defined as the moment of the
distribution function as 
\begin{equation}
\rho =m\int fd\mathbf{c}dI\text{, }\rho v_{i}=m\int fc_{i}d\mathbf{c}dI,
\label{Moments density and velocity}
\end{equation}%
\begin{equation}
\rho u=m\int f\left( \frac{C^{2}}{2}+I\right) d\mathbf{c}dI , \label{moments energy}
\end{equation}%
where $m$ is the molecular mass, and $\mathbf{C} =\mathbf{c}-\mathbf{v}$ is the
peculiar velocity with respect to the macroscopic velocity $\textbf{v}$ (or $v_{i}$ in tensorial notation). The differential velocity vector
$d\mathbf{c}$ implies integration over all three components of velocity space. Furthermore, the internal energy $u$, can be divided into
the translational part $u^{tr}$ and the part due to the internal degrees of
freedom $u^{in}$, as%
\begin{eqnarray}
\rho u^{tr}&:=&\frac{3}{2}\rho \theta ^{tr} =m\int f\frac{C^{2}}{2}d\mathbf{c}dI,
\label{moments translational energy} \\
\rho u^{in}&:=&\frac{\delta }{2}\rho \theta ^{in} =m\int fId\mathbf{c}dI\text{.}
\label{moments internal energy}
\end{eqnarray}%
Here, subscripts ``\textit{tr}'' and ``\textit{in}'' denote the translational and internal parts, while the parameter $\delta$
 is the number of internal degrees of freedom.
Conventionally, we define the translational temperature $\theta^{tr}$ and the internal temperature $\theta^{in}$ in energy units. As a result, the thermodynamic temperature $\theta$ (defined as $\theta:= RT$; $R$ being the gas constant) can be expressed as follows:
\begin{equation}
\frac{3+\delta }{2}\theta =\frac{3}{2}\theta ^{tr}+\frac{\delta }{2}\theta ^{in}\text{.}  \label{temperature}
\end{equation}
When the system is in equilibrium, the three temperatures are equal: $\theta^{tr} = \theta^{in} = \theta$. However, in a non-equilibrium state, these temperatures can differ. To quantify the non-equilibrium part of the temperature, we introduce the dynamic temperature denoted as $\vartheta = \theta^{tr} - \theta$. In this context, the dynamic pressure in the gas can be expressed as $\Pi = \rho \vartheta$.
\\
The pressure tensor $p
_{ij}$ is defines as follows
\begin{equation}
p _{ij}=m\int fC_{i}C_{j}d\mathbf{c}dI,
\label{moments translational temperature tensor}
\end{equation}%
so that its trace $p _{kk}=3 \rho \theta ^{tr}$. Furthermore, the pressure tensor is expressed in terms of its trace and traceless part
as
 \begin{equation}
    p_{ij} =(p+\Pi)\delta_{ij}+\sigma_{\langle ij\rangle}\text{,}
 \end{equation}
 where $\delta_{ij}$ is the Kronecker delta function, the angular brackets around indices represent the symmetric and traceless part of a tensor, $p$ is the equilibrium pressure, $\sigma_{ij}$ is the viscous stress tensor and $\Pi$ is the dynamic pressure (or nonequilibrium pressure). 
 % The internal energy density of the system can be divided into two components: translational energy $(\rho u^{tr})$ and internal energy $(\rho u^{in})$, represented as $\rho u = \rho u^{tr} + \rho u^{in}$. From this, we can introduce the concepts of translational temperature and internal temperature, defined by equations \eqref{moments translational energy} and \eqref{moments internal energy}, respectively. Building upon these temperature definitions, 
 We can further define the translational heat flux and internal heat flux as
 % The symmetric traceless part $\sigma_{\langle ij\rangle}$ is called the shear stress tensor.
\begin{eqnarray}
q_{k}^{in} &:=&m\int fIC_{k}d\mathbf{c}dI ,\\
q_{k}^{tr} &:=&m\int f\frac{C^{2}}{2}C_{k}d\mathbf{c}dI
\label{heat flux both}
\end{eqnarray}
Furthermore, the total heat flux $q_k$ and the heat flux difference $Q_k$ are introduced as moments of the distribution functions. They are defined as follows:
\begin{eqnarray}
q_{k} &:=&m\int f\left( \frac{C^{2}}{2}+I\right)
C_{k}d\mathbf{c}dI=q_{k}^{tr}+q_{k}^{in} ,\\
 Q_{k} &:=&{q^{tr}_{k}}-\frac{5\theta+3\vartheta}{\delta\theta-3\vartheta}q^{in}_{k},\label{heat flux difference}
\end{eqnarray}%
%here particular choice of $Q_{k}$ motivated from order of magnitude approach by Behnam Rahimi \cite{rahimi2016macroscopic} where it was shown this term is order of $Kn^{1+\alpha}$, here $\alpha$ is a magnifying parameter, with $0<\alpha<1$.
The idea behind selecting a particular expression for $Q_{k}$ was motivated by the order-of-magnitude approach proposed by Rahimi \& Struchtrup \cite{rahimi2016macroscopic}. This approach revealed that the magnitude of this term can be approximated as $O(Kn^{1+\alpha})$, where $\alpha$ is a magnification parameter falling within the range of 0 to 1. Furthermore, opting for a specific combination of $q^{tr}_k$ and $q^{in}_k$ in a linear fashion results in a convenient expression for the entropy flux, which will be further discussed in Section \S\ref{The second law of thermodynamics}. %Furthermore with this particular choice of linear combination of $q^{tr}_k$ and $q^{in}_k$ results into convenient from of the entropy flux which we discuss later in section 2.

\subsection{The conservation laws and  extended balance equations}
The conservation laws are obtained from the Boltzmann equation (\ref%
{Boltzman equation}), by multiplying it with the collision invariants (\ref%
{collision invariants}) and integrating, to get
\begin{subequations}
\label{Conservations laws}
\begin{eqnarray}
\frac{D\rho }{Dt}+\rho \frac{\partial v_{k}}{\partial x_{k}} &=&0\text{,}
\label{mass  conservation} \\
\rho \frac{Dv_{i}}{Dt}+\frac{\partial p _{ik}}{\partial x_{k}} &=&0%
\text{,}  \label{momentum conservation} \\
\rho \frac{D\frac{3+\delta }{2}\theta }{Dt}+p _{ik}\frac{\partial
v_{i}}{\partial x_{k}}+\frac{\partial q_{k}}{\partial x_{k}} &=&0\text{.}
\label{energy conservation}
\end{eqnarray}%
Here, $\frac{D}{D t} =\frac{\partial}{\partial t}+v_{k}\frac{\partial}{\partial x_k}$ is the convective time derivative. We also get the balance equations for the translational temperature
and internal temperature as follows 
\end{subequations}
\begin{eqnarray}
\rho \frac{3}{2}\frac{D\theta ^{tr}}{Dt}+p _{ik}\frac{\partial
v_{i}}{\partial x_{k}}+\frac{\partial q_{k}^{tr}}{\partial x_{k}} &=&%
\mathcal{P}^{0,0}  ,\label{translational temperature} \\
\rho \frac{\delta }{2}\frac{D\theta ^{in}}{Dt}+\frac{\partial q_{k}^{in}}{%
\partial x_{k}} &=&\mathcal{P}^{0,1}.  \label{internal temperature}
\end{eqnarray}%
The production terms $\mathcal{P}
^{0,0}$ and $\mathcal{P}^{0,1}$ are obtained from the Bolzmann collision operator, which entails that
\begin{equation*}
\text{ }\mathcal{P}^{0,0}:=m\int \mathcal{S}\frac{C^2}{2}d\mathbf{c}dI\text{, and }%
\mathcal{P}^{0,1}:=m\int \mathcal{S}Id\mathbf{c}dI\text{,}
\end{equation*}%
therefore $\mathcal{P}^{0,0}=-\mathcal{P}^{0,1}$, since $\frac{C^{2}}{2%
}+I$ is collison invarient.
Further, the balance law for dynamic temperature is written as 
\begin{eqnarray}
 \frac{3 \rho }{2}\frac{D\vartheta}{Dt}+\frac{\delta}{3+\delta}p _{ik}\frac{\partial
v_{i}}{\partial x_{k}}-\frac{3}{3+\delta}\frac{\partial q_{k}}{\partial x_{k}}+\frac{\partial q_{k}^{tr}}{\partial x_{k}} =%
\mathcal{P}^{0,0} . \label{dynamic temperature} 
\end{eqnarray}
In terms of total heat and heat difference flux, the last equation can also be written as
\begin{multline}
\frac{3 \rho }{2}\frac{D\vartheta}{Dt}+\frac{\delta}{3+\delta}p _{ik}\frac{\partial
v_{i}}{\partial x_{k}}+\frac{2}{5+\delta}\left(\frac{\delta}{3+\delta}+\frac{3}{2}\frac{\vartheta}{\theta}\right)\frac{\partial q_{k}}{\partial x_{k}}\\
+\frac{3}{5+\delta}q_{k}\frac{\partial }{\partial x_{k}}\left(\frac{\vartheta}{\theta}\right)
+\frac{\delta}{5+\delta}\left(1-\frac{3}{\delta}\frac{\vartheta}{\theta}\right)\frac{\partial Q_{k}}{\partial x_{k}}\\
-\frac{3}{5+\delta}Q_{k}\frac{\partial }{\partial x_{k}}\left(\frac{\vartheta}{\theta}\right) =
\mathcal{P}^{0,0}. \label{dynamic temperature 2} 
\end{multline}

% This file may be formatted in both the \texttt{preprint} (the default) and
% \texttt{reprint} styles; the latter format may be used to 
% mimic final journal output. Either format may be used for submission
% purposes; however, for peer review and production, AIP will format the
% article using the \texttt{preprint} class option. Hence, it is
% essential that authors check that their manuscripts format acceptably
% under \texttt{preprint}. Manuscripts submitted to AIP that do not
% format correctly under the \texttt{preprint} option may be delayed in
% both the editorial and production processes.
% \\
% The \texttt{widetext} environment will make the text the width of the
% full page, as on page~\pageref{eq:wideeq}. (Note the use the
% \verb+\pageref{#1}+ to get the page number right automatically.) The
% width-changing commands only take effect in \texttt{twocolumn}
% formatting. It has no effect if \texttt{preprint} formatting is chosen
% instead.
The governing equations (\ref{Conservations laws})--(\ref{translational temperature}) describe flow behavior but require additional closure models to relate unknown fluxes \{$\sigma_{ij},q^{tr}_{k},q^{in}_{k}$\} to known quantities $\{\rho,v_i,\theta^{tr},\theta^{in}\}$. 
\section{\label{The second law of thermodynamics}The second law of thermodynamics}
This section focuses on obtaining the non-equilibrium distribution function $f_{\mathrm{6}}$ by applying the maximum entropy principle. Additionally, the second law of thermodynamics is proven, and constitutive relations are derived. The constitutive relations are functional relations between the dependent fields \{$\sigma_{ij},q^{tr}_{k},q^{in}_{k}$\} and the independent fields $\{\rho,v_i,\theta^{tr},\theta^{in}\}$ that is we expressed independent field variables as functional relations of dependent field variables.\\
The second law of thermodynamics asserts that a physical system in equilibrium 
has maximal entropy among all states with the same energy.
The second law of thermodynamics, which we now introduce in the form of the entropy principle. The entropy principle will always be exploited for supply-free bodies, i.e., there are no body forces and no radiation, nor is there a supply of entropy. There exists a specific entropy of the gas $s$, the non-convective entropy flux $h_{k}$, and entropy production density $\sum$, which obey a balance law.

The entropy density is defined by the relation
\begin{equation}
   \rho s= -k_{b}\int f\ln \frac{f}{f_{0}}dCdI, \label{entropy equation}
\end{equation}
and the entropy law
\begin{equation}
    \rho \frac{Ds}{Dt}+\frac{\partial h_{k}}{\partial x_{k}}=\Sigma~,\label{entropy law 1}
\end{equation}
where $k_{b}$ is the Boltzmann constant, $\Sigma$ the entropy production  and $f_{0}=I^{(\frac{\delta}{2}-1)}$. The second law of thermodynamics requires entropy production density non-negative i.e., $\Sigma \geq 0$. Now, any process satisfying the second law represents a so-called physically admissible process. The density of entropy production is non-negative for all thermodynamic processes, i.e. for all solutions of the field equations. Thus, the entropy inequality $\left(\Sigma \geq 0\right)$ holds for all thermodynamic processes.
First of all, we determine the phase density $ f_6$ that maximizes the entropy density $\rho s$ under the constraints of fixed mass density$\rho$, momentum density $\rho v_i$ and internal energy density $\rho u$ (both parts). After that, we introduce it into the expressions \eqref{entropy equation} 
to obtain the entropy density.
\begin{theorem}
The maximum entropy distribution function which maximizes the entropy (\ref{entropy equation}) under the constraints (\ref%
{Moments density and velocity}), (\ref{moments translational energy}) and (%
\ref{moments internal energy}) takes the following form \cite{frezzotti2007numerical}:
\begin{equation*}
f_{\mathrm{6}}=\underset{\text{Maxwellian}}{\frac{\rho }{m}\underbrace{\frac{%
1}{\left(2\pi \theta ^{tr}\right)^{3/2}} e^{-\frac{C^{2}}{2\theta _{tr}}}}}\underset{%
\text{Gamma}}{\underbrace{\frac{1}{\Gamma \left( \frac{\delta }{2}\right) }%
\frac{1}{I}\left( \frac{I}{\theta ^{in}}\right) ^{\delta /2}e^{-\frac{I}{%
\theta ^{in}}}}} .
\end{equation*}%
%Proof:%
\begin{proof}
See Appendix.
\end{proof}
\end{theorem}
Therefore, the entropy density for the 6-moment system is given by the relation
\begin{equation}
\rho s =-k_{b}\int f_{6}\ln \left( \frac{f_{6}}{f_{0}}\right) d\mathbf{c}dI=\rho
\left\{\frac{3}{2} \ln  \theta ^{tr}+\frac{\delta }{2}\ln \theta ^{in}-\ln \rho \right\}. \label{entropy density}
\end{equation}% 
Moreover, The extended Gibbs' relation is given by:
\begin{equation}
    \rho ds=\frac{3}{2\theta ^{tr}}\rho d\theta ^{tr}+\frac{\delta }{2\theta
^{in}}\rho d\theta ^{in}-d\rho. \label{extended gibbs relation}
\end{equation}
The proof for both equations \eqref{entropy density} and \eqref{extended gibbs relation} is simple and straightforward.  Gibbs' relation can be obtained by differentiating equation \eqref{entropy density}.
Now, using the extended Gibbs equation (\ref{extended gibbs relation}),
the time rate of change of the entropy of a material element is given as
\begin{equation}
    \rho \frac{Ds}{Dt}=\frac{3}{2\theta ^{tr}}\rho \frac{D\theta ^{tr}}{Dt}+%
\frac{\delta }{2\theta ^{in}}\rho \frac{D\theta ^{in}}{Dt}-\frac{D\rho }{Dt}.
\label{entropy balance 23}
\end{equation}
Using equations (\ref{mass  conservation}), (\ref{translational temperature}) and (\ref{internal temperature}) in (\ref{entropy balance 23}), we get
\begin{eqnarray}
    \rho \frac{Ds}{Dt}+\frac{\partial \left( \frac{q_{k}^{tr}}{\theta ^{tr}}+%
\frac{q_{k}^{in}}{\theta ^{in}}\right) }{\partial x_{k}} =&&-\frac{1}{\theta
^{tr}}\sigma _{ik}\frac{\partial v_{i}}{\partial x_{k}}+q_{k}^{tr}\frac{%
\partial \left( \frac{1}{\theta ^{tr}}\right) }{\partial x_{k}}\\&&
+q_{k}^{in}%
\frac{\partial \left( \frac{1}{\theta ^{in}}\right) }{\partial x_{k}}\nonumber+\left(
\frac{1}{\theta ^{in}}-\frac{1}{\theta ^{tr}}\right) \mathcal{P}^{0,1},
\label{entropy law a}
\end{eqnarray}
Comparing the above equation with the entropy law (\ref{entropy law 1}), we identify the non-convective entropy flux $h_k$ as follows
\begin{equation}
    h_k= \frac{q_{k}^{tr}}{\theta ^{tr}}+%
\frac{q_{k}^{in}}{\theta ^{in}}\text{.}
\label{entropy flux 25}
\end{equation}
Clearly, when the system is in equilibrium ($\theta ^{tr}=\theta ^{in}=\theta$), the three temperatures are equal, and the above expression simplifies to the classical entropy flux expression, i.e., $h_k=q_k/\theta$. \\
By introducing the variables $q_k$, $Q_k$, $\theta$, and $\vartheta$, equation (\ref{entropy flux 25}) can be rewritten as follows:
\begin{equation}
    h_k=\frac{q_{k}}{\theta}-\frac{2}{5+\delta}\frac{\vartheta}{\theta(\theta+\vartheta)}q_{k}-\frac{3+\delta}{5+\delta}\frac{\vartheta}{\theta(\theta+\vartheta)}Q_k. \label{entropy flux}
\end{equation}
Indeed, in equation \eqref{heat flux difference}, we have introduced $Q_k$ so that entropy flux can be conveniently decomposed into three components. The first contribution to the entropy flux is classical, the second one stems from coupled constitutive relations \cite{pavic2013maximum}, whereas the last is the contribution of higher order. Furthermore 
%it can be shown that $h_k$ via order of magnitude 
via order of magnitude analysis, it can be shown that %analysis has been shown that
first term is of the order $O(Kn^{1})$ and while second term is order $O(Kn^{1+\alpha})$ whilest third term is order of $O(Kn^{1+2\alpha})$.

% \begin{eqnarray}
%     \rho \frac{Ds}{Dt}+\frac{\partial \left[ \frac{q_{k}}{\theta}-\left(\frac{2}{5+\delta}\right)\left(\frac{\vartheta}{\theta(\theta+\vartheta)}\right)q_{k}-\left(\frac{3+\delta}{5+\delta}\right)\left(\frac{\vartheta}{\theta(\theta+\vartheta)}\right)Q_k\right] }{\partial x_{k}} =&&\nonumber\\
%     -\frac{1}{\theta
% ^{tr}}\sigma _{ik}\frac{\partial v_{i}}{\partial x_{k}}-\frac{1}{{\theta ^{tr}}^2} q_{k}^{tr}\frac{%
% \partial \theta ^{tr}}{\partial x_{k}}-\frac{1}{{\theta ^{in}}^2}q_{k}^{in}%
% \frac{\partial \theta ^{in}}{\partial x_{k}}+\left[ 
% \frac{1}{\theta ^{in}}-\frac{1}{\theta ^{tr}}\right] \mathcal{P}^{0,1}.\nonumber\\
% \label{entropy law 2}
% \end{eqnarray}

Again, comparing equation (\ref{entropy law a}) with the entropy law (\ref{entropy law 1}), we get the entropy production rate, as
\begin{eqnarray}
    \Sigma=&&-\frac{1}{\theta
^{tr}}\sigma _{ik}\frac{\partial v_{i}}{\partial x_{k}}-\frac{1}{{\theta ^{tr}}^2} q_{k}^{tr}\frac{%
\partial \theta ^{tr}}{\partial x_{k}}-\frac{1}{{\theta ^{in}}^2}q_{k}^{in}
\frac{\partial \theta ^{in}}{\partial x_{k}}\nonumber\\&&
+\left( 
\frac{1}{\theta ^{in}}-\frac{1}{\theta ^{tr}}\right) \mathcal{P}^{0,1}\text{,}
\end{eqnarray}
which is in the bilinear form for flux and gradients. The second law of thermodynamics requires that the entropy production rate must be positive, i.e., $\Sigma \geq 0$. To ensure the positivity of the entropy production rate $\Sigma$, it is sufficient to assume relations of the form:
\begin{subequations}
%\label{constitutive relations}
\begin{eqnarray}
\sigma _{ik} &=&-2\mu \frac{\partial v_{\langle i}}{\partial x_{k\rangle }}\text{,} 
\label{stress tensor constitutive relation 1} 
\\
q^{tr}_{k}&=&-\zeta_{11} \frac{1}{{\theta ^{tr}}^2}\frac{\partial\theta^{tr}}{\partial x_{k}}-\zeta_{12}\frac{1}{{\theta ^{in}}^2} \frac{\partial\theta^{in}}{\partial x_{k}}\text{, and}
\label{translational heat flux constitutive relation 1}
\\
q^{in}_{k}&=&-\zeta_{12}\frac{1}{{\theta ^{tr}}^2} \frac{\partial\theta^{tr}}{\partial x_{k}}-\zeta_{22} \frac{1}{{\theta ^{in}}^2}\frac{\partial\theta^{in}}{\partial x_{k}} .
\label{Internal heat flux constitutive relation 1}
\end{eqnarray}
\end{subequations}
where $\mu \geq 0$ is viscosity of the gas and the matrices
\begin{equation}
    \eta=\begin{bmatrix}
        \zeta_{11} & \zeta_{12}\\ 
         \zeta_{12} & \zeta_{22} 
     \end{bmatrix}
\end{equation}
 is  a symmetric non-negative definite matrix.\\ 
One can express the total heat flux ($q_k$) and the heat difference ($Q_k$), using (\ref{translational heat flux constitutive relation 1})--(\ref{Internal heat flux constitutive relation 1})
\begin{eqnarray}
    q_k=&&-\left[\frac{\zeta_{11}+\zeta_{12}}{(\theta+\vartheta)^2}+\frac{(\zeta_{12}+\zeta_{22})\delta^2}{(\delta\theta-3\vartheta)^2}\right] \frac{\partial \theta}{\partial x_{k}}\nonumber\\&&
    -\left[\frac{\zeta_{11}+\zeta_{12}}{(\theta+\vartheta)^2}-\frac{3(\zeta_{12}+\zeta_{22})\delta}{(\delta\theta-3\vartheta)^2}\right] \frac{\partial\vartheta}{\partial x_{k}} ~~~\text{, and}\label{total heat flux con relation}
\end{eqnarray}

\begin{multline}
    Q_{k}=-\left[\frac{\zeta_{11}}{(\theta+\vartheta)^2}-\frac{\zeta_{22}(5\theta+3\vartheta)\delta^2}{(\delta\theta-3\vartheta)^3}\right.\\
    \left.-\frac{\zeta_{12}\{-\delta^2(\theta+\vartheta)^2+\delta\theta(5\theta+3\vartheta)-3\vartheta(5\theta+3\vartheta)\}}{(\theta+\vartheta)^2(\delta\theta-3\vartheta)^2}\right] \frac{\partial\theta}{\partial x_{k}}\\
    -\left[\frac{\zeta_{11}}{(\theta+\vartheta)^2}+\frac{3\zeta_{22}(5\theta+3\vartheta)\delta}{(\delta\theta-3\vartheta)^3}\right.\\
    \left.-\frac{\zeta_{12}\{8\delta\theta^2+3(-5+3\delta)\theta\vartheta+3(-3+\delta)\vartheta^2\}}{(\theta+\vartheta)^2(\delta\theta-3\vartheta)^2}\right] \frac{\partial\vartheta}{\partial x_{k}}\text{.\qquad }
    \label{total heat diff flux con relation}
\end{multline}
Furthermore, one can show that the last term in (\ref{entropy law a}) is positive, if we take $\mathcal{P}^{0,1}$ to be proportional to $\vartheta$, as
\begin{equation}
\begin{split}
\left[ \frac{1}{\theta ^{in}}%
-\frac{1}{\theta ^{tr}}\right] \mathcal{P}^{0,1} & =\left[ \frac{\theta ^{tr}-\theta ^{in}}{\theta ^{tr}\theta ^{in}}\right]\frac{\rho \vartheta }{\tau _{int}} 
\\
& =\frac{\rho}{\tau _{int}}\left[ \frac{\theta ^{tr}-\theta ^{in}}{\theta ^{tr}\theta ^{in}}\right]\left(\theta ^{tr}-\theta\right)
 \\
 &=\frac{\rho \delta}{\tau _{int} (3+\delta)}\left[ \frac{(\theta ^{tr}-\theta ^{in})^2}{\theta ^{tr}\theta ^{in}}\right]\\
 &\geq 0\text{.}
 \end{split}
\end{equation}
where $\tau _{int}>0$ is the relaxation time for $\vartheta$. This completes the proof of H-theorem (second law of thermodynamics) for the two-temperature model.
% Now comparing equations (\ref{entropy law 1}) and (\ref{entropy law 2}) we get entropy flux $h_k$ as follows
% \begin{equation}
%     h_k=\frac{q_{k}}{\theta}-\left(\frac{2}{5+\delta}\right)\left(\frac{\vartheta}{\theta(\theta+\vartheta)}\right)q_{k}-\left(\frac{3+\delta}{5+\delta}\right)\left(\frac{\vartheta}{\theta(\theta+\vartheta)}\right)Q_k. \label{entropy flux}
% \end{equation}
%%%%%%%%%%%%%%%%%%%%%%%%%%%%%%%%%%%%%%%%%%%%%%%
\subsection{Determination  of phenomenological coefficients}
In this section, we find the value of 
arbitrary non-negative coefficients $\mu$, $\zeta_{11}$, $\zeta_{22}$, $\zeta_{12}$ and production terms  $\mathcal{P}^{0,1}$ via comparison with different models from literature. The values for these phenomenological coefficients and production terms may vary with different collision models. Here, we consider four models which give different values. 
\subsubsection{Comparison with Marques and Kremer \cite{marques1993spectral} for $\delta=3$ (Model 1)}
Marques and Kremer \cite{marques1993spectral} introduced a hydrodynamical model that includes a relaxation equation for the dynamic pressure, which is the non-equilibrium component of the pressure, in addition to the conventional hydrodynamic equations. We compare proposed model balance equations and constitutive relations with hydrodynamical model\cite{marques1993spectral} equations; we get non-negative coefficients and production terms as

\begin{subequations}
\label{eq:whole}
\begin{eqnarray}
 \mu&=&\frac{15}{8 a^2}\left(\frac{k_{b}\theta m}{\pi}\right)^{1/2}\frac{(\kappa+1)^2}{(13\kappa+6)}\text{,}
 %\label{subeq:1}
 \\
% \end{eqnarray}
% \begin{eqnarray}
 \zeta_{11}&=& \frac{15\mu(6+25\kappa+38\kappa^{2}+26\kappa^3)}{24+150\kappa+202\kappa^{2}+204\kappa^3}\text{,}
%\label{subeq:2}
\\
% \end{eqnarray}
% \begin{eqnarray}
    \zeta_{12}&=& \frac{15\kappa(6+13\kappa)\mu}{24+150\kappa+202\kappa^{2}+204\kappa^3}\text{,}
 \\
% \end{eqnarray}
% \begin{eqnarray}
     \zeta_{22}&=& \frac{9(24+154\kappa+221\kappa^2)\mu}{10(12+75\kappa+101\kappa^{2}+102\kappa^3)}\text{,}
\\     
% \end{eqnarray}
% \begin{eqnarray}
    \mathcal{P}^{0,1}&=&\frac{\delta\rho  }{(3+\delta)\tau _{int}}\left(\theta^{tr}-\theta ^{in}\right)\text{,}
\end{eqnarray}
\end{subequations}
where $\mu$ is the shear viscosity, $\zeta_{11}$ the thermal conductivity for the translational temperature, $\zeta_{22}$ the thermal conductivity for the internal temperature,
\begin{equation}
    \frac{1}{\tau _{int}}=\frac{32}{3}a^{2}\frac{\rho }{m}\sqrt{\pi \theta}\frac{%
\kappa }{\left( 1+\kappa \right) ^{2}},
\end{equation}
is the relaxation frequency of the dynamic pressure, $a$ and $\kappa=\frac{4 I}{m a^2}$ are the diameter 
and the dimensionless moment of inertia of the spherical molecule respectively \cite{gaio1991kinetic}. The range of $\kappa \in \left[0,2/3\right]$, with $0$ denoting no mass distribution on the surface of the molecule and $2/3$ denoting uniform mass distribution.
%%%%%%%%%%%%%%%%%%%%%%%%%%%%%%%%%%%%%
\subsubsection{Comparison with Aoki et al. \cite{aoki2020two} (Model 2)}
Aoki et al. \cite{aoki2020two} developed the Navier-Stokes equations for a polyatomic gas that exhibits slow relaxation of its internal modes, using the ellipsoidal-statistical model of the Boltzmann equation proposed by Andries et al. \cite{andries2000gaussian}. This model considers two temperatures, translational and internal, and Aoki et al. used the Chapman-Enskog method to derive the equations. On comparing the balance equations and constitutive relations of the proposed model with the equations of the Aoki et al. model, we find the non-negative coefficients and production terms as 
\begin{subequations}
%\label{eq:whole}
\begin{eqnarray}
 \mu&=& \frac{1}{1-\nu} \frac{\theta ^{tr}}{2\mathcal{A}_{c}(T)}\text{,}
 %\label{subeq:1}
\\
 \zeta_{11}&=& \frac{5\mu}{2}\frac{(\theta ^{tr})^{3}}{\mathcal{A}_{c}(T)}\text{,}
%\label{subeq:2}
\\
    \zeta_{12}&=& ~0\text{,}
\\
    \zeta_{22}&=& \frac{\delta\mu}{2}\frac{(\theta ^{in})^{2}\theta ^{tr}}{\mathcal{A}_{c}(T)}\text{,}
\\
    \mathcal{P}^{0,1}&=& \frac{3\delta}{2(3+\delta)}\theta_{1}\mathcal{A}_{c}(T)\rho^{2}\left(\theta^{tr}-\theta ^{in}\right)\text{,}
\end{eqnarray}
\end{subequations}
where $\nu \in [-\frac{1}{2},1)$, $\theta_{1} \in (0, 1]$ are the constants that adjust the
Prandtl number and the bulk viscosity. In addition, $\mathcal{A}_{c}(T)$ is a
function of $T$ such that $\rho\mathcal{A}_{c}(T)$ is the collision frequency of
the gas molecules.
%%%%%%%%%%%%%%%%%%%%%%%%%%%%%%%%%%%%%%%%%%%%%
\subsubsection{Comparison with Djordji\'{c} et al.\cite{djordjic2023boltzmann} (Model 3)}
The macroscopic system of 17-moment equations was derived by Djordji\'{c} et al. \cite{djordjic2023boltzmann} from the Boltzmann equation after proposing a collision kernel for polyatomic gases with continuous internal energy. These equations were then used to calculate important transport properties, including shear viscosity, Prandtl number, and the ratio of bulk to shear viscosities. After that, the proposed collision kernel was then employed to compute these transport properties in the polytropic regime for various polyatomic gases. By examining the balance equations and constitutive relations of the proposed model and comparing them to those of the reduced 17-moment system, we identified both the production terms and  non-negative coefficients as
\begin{subequations}
%\label{eq:whole}
\begin{eqnarray}
 \mu&=& \frac{p}{P^{0}_{\sigma}}\text{,}
 %\label{subeq:1}
\\
 \zeta_{11}&=& \frac{5\mu (\theta ^{tr})^{2}}{2}\frac{P^{0}_{s}}{(P^{0}_{q}P^{0}_{s}-P^{1}_{q}P^{1}_{s})}\text{,}
%\label{subeq:2}
\\
    \zeta_{12}&=& -\frac{\delta \mu(\theta ^{in})^{2}}{2}\frac{P^{1}_{q}}{(P^{0}_{q}P^{0}_{s}-P^{1}_{q}P^{1}_{s})}\text{,}
\\
    \zeta_{21}&=&-\frac{5 \mu(\theta ^{tr})^{2}}{2}\frac{P^{1}_{s}}{(P^{0}_{q}P^{0}_{s}-P^{1}_{q}P^{1}_{s})}\text{,}
\\
    \zeta_{22}&=& \frac{\delta \mu(\theta ^{in})^{2}}{2}\frac{P^{0}_{q}}{(P^{0}_{q}P^{0}_{s}-P^{1}_{q}P^{1}_{s})}\text{,}
\\
    \mathcal{P}^{0,1}&=& \frac{3\delta\rho}{2(3+\delta)}(\theta^{tr}-\theta^{in})P^0_{\Pi}\text{,}
\end{eqnarray}
\end{subequations}
here $P^{0}_{q},P^{1}_{q},P^{0}_{s},P^{1}_{s},P^{0}_{\Pi} \text{ and } P^{0}_{\sigma}$ are constants that have different values for different gases. These constants can be directly computed using the Mathematica\textregistered~  code in \cite{djordjic2021explicit}. 
%%%%%%%%%%%%%%%%%%%%%%%%%%%%%%%%%%%%%%%%%%%%%%%
\subsubsection{Comparison with Rahimi and Struchtrup \cite{rahimi2016macroscopic} (Model 4)}
By comparing the balance equations and constitutive relations of our proposed model with the high-order macroscopic model developed by Rahimi and Struchtrup \cite{rahimi2016macroscopic}, we identify
\begin{equation}
  \renewcommand{\arraystretch}{1.2}% To spread out the equations
  \left.\begin{array}{r@{\;}l}
     \zeta_{11}=&\frac{5 \lambda}{5+\delta}(\theta ^{tr})^{2}\text{,} \\
     \zeta_{22}=&\frac{\delta \lambda}{5+\delta}(\theta ^{in})^{2}\text{,} \\ \text{and}~~~~~~~~~~~~~~~~~~~
    \zeta_{12}=&0 \text{,}
  \end{array}\right\} %\label{tensor eq1}
\end{equation}
where $\lambda$ is the thermal conductivity of the gas.
%%%%%%%%%%%%%%%%%%%%%%%%%%%%%%%%%%%%%%%%
\section{Linearized and dimensionless equations}
\label{Linearized and dimensionless equations}
This section considers dimensionless and linearized equations by introducing small perturbations from their values in a reference rest state characterized by a constant pressure $p_0$ and a constant temperature $\theta_0$. The relationships between the field variables and their dimensionless deviations (denoted with hat symbols) from the reference rest state are expressed as:
\begin{eqnarray}
    p&=&p_{0}(1+\hat{p})\text{,}~ \theta=\theta_{0}(1+\hat{\theta}) \text{,} ~v_{i}=\sqrt{\theta_{0}}\hat{v}_{i}\text{,}~\sigma_{ij}=p_{0}\hat{\sigma}_{ij}\text{,}\nonumber\\
q_{i}&=&p_{0}\sqrt{\theta_{0}}\hat{q}_{i}~~
    \text{and}~~ x_i=L\hat{x}_i,
    \label{dimensionless variable}
\end{eqnarray}
where $L$ is a characteristic length scale. The linearized conservation laws and balance equations are given by
\begin{subequations}
\label{linear equation 1}
\begin{eqnarray}
  \frac{\partial \hat{\rho}}{\partial \hat{t}}+ \frac{\partial \hat{v}_i}{\partial \hat{x}_i}&=&0\text{,}
 %\label{subeq:1}
\\
 \frac{\partial \hat{v}_i}{\partial \hat{t}}+\frac{\partial \hat{\rho}}{\partial\hat{x}_i}+\frac{\partial \hat{\theta}^{tr}}{\partial \hat{x}_i}+\frac{\partial \hat{\sigma}_{ij}}{\partial \hat{x}_j}&=&0\text{,}
%\label{subeq:2}
\\
    \frac{3+\delta}{2}\frac{\partial \hat{\theta}}{\partial \hat{t}}+\frac{\partial \hat{v}_i}{\partial \hat{x}_i} + \frac{\partial \hat{q}_{i}}{\partial \hat{x}_i}&=&0\text{,}
\\
   \frac{3}{2}\frac{\partial \hat{\theta}^{tr}}{\partial \hat{t}}+\frac{\partial \hat{v}_i}{\partial \hat{x}_i} +\frac{\partial \hat{q}_{i}^{tr}}{\partial \hat{x}_i}&=&\mathcal{P}^{0,0}\text{,}
\\
     \frac{\delta}{2}\frac{\partial \hat{\theta}^{in}}{\partial \hat{t}}+ \frac{\partial \hat{q}_{i}^{in}}{\partial \hat{x}_i}&=&\mathcal{P}^{0,1}\text{,}
\end{eqnarray}
\end{subequations}
and the linearized stress, translational, and internal heat fluxes are specified as
\begin{subequations}
\label{linear equation 2}
\begin{eqnarray}
  \hat{\sigma}_{ij} &=&-2 Kn \frac{\partial \hat{v}_{\langle i}}{\partial \hat{x}_{j\rangle }}\text{,}
 %\label{subeq:1}
\\
 \hat{q}_{i}^{tr}&=&-\zeta_{11}\frac{\partial\hat{\theta}^{tr}}{\partial \hat{x}_i}-\zeta_{12}\frac{\partial\hat{\theta}^{in}}{\partial \hat{x}_i}\text{,~~and}
%\label{subeq:2}
\\
   \hat{q}_{i}^{in}&=&-\zeta_{21} \frac{\partial \hat{\theta}^{tr}}{\partial \hat{x}_i}-\zeta_{22}\frac{\partial\hat{\theta}^{in}}{\partial \hat{x}_i}\text{.}
\end{eqnarray}
\end{subequations}
Next linearized the total heat flux $ q_k$ and heat flux difference $Q_k$
\begin{equation}
    \hat{q}_i=-(\zeta_{11}+2\zeta_{12}+\zeta_{22}) \frac{\partial \hat{\theta}}{\partial \hat{x}_i}-\left((\zeta_{11}+\zeta_{12})-\frac{3}{\delta}(\zeta_{12}+\zeta_{22})\right) \frac{\partial\hat{\vartheta}}{\partial \hat{x}_i} \text{,}
    \label{total heat flux con relation linear}
\end{equation}
\text{and}
\begin{eqnarray}
    \hat{Q}_{i}=&&-\left((\zeta_{11}-\frac{5}{\delta}\zeta_{12})+(\zeta_{12}-\frac{5}{\delta}\zeta_{22})\right) \frac{\partial\hat{\theta}}{\partial \hat{x}_i}\nonumber\\&&
    -\left((\zeta_{11}-\frac{5}{\delta}\zeta_{12})-\frac{3}{\delta}(\zeta_{12}-\frac{5}{\delta}\zeta_{22})\right) \frac{\partial\hat{\vartheta}}{\partial \hat{x}_i}\text{.\qquad }
    \label{total heat diff flux con relation linear}
\end{eqnarray}
In (\ref{linear equation 2}), the Knudsen number appears as the scaled viscosity $Kn=\mu_{0}\sqrt{\theta_{0}}/(p_{0}L)$.
%%%%%%%%%%%%%%%%%%%%%%%%%%%%%%%%%%%%
\section{\label{Reduced model}Reduced model}
In this section, we have introduced a reduced model derived from the present two-temperature model. In this simplified model, we assume that the heat fluxes (represented by $Q_{k}$) are negligible, resulting in the following relationships between the phenomenological coefficients.
\begin{equation}
    \zeta_{12}=\frac{\delta\theta-3\vartheta}{5\theta+3\vartheta}\zeta_{11}\text{~~~~~and\qquad }\zeta_{22}=\frac{(\delta\theta-3\vartheta)^2}{(5\theta+3\vartheta)^2}\zeta_{11}.
\end{equation}
Now write linearized reduced model equations in six field variable mass density $\rho$, velocity $v_i$, temperature $\theta$, and the dynamic temperature $\vartheta$ 
\begin{subequations}
\label{reduced linear equation 1}
\begin{eqnarray}
  \frac{\partial \rho}{\partial t}+ \frac{\partial v_i}{\partial x_i}&=&0\text{,}
 %\label{subeq:1}
\\
  \frac{\partial v_i}{\partial t}+\frac{\partial \rho}{\partial x_i}+\frac{\partial \theta}{\partial x_i}+\frac{\partial \vartheta}{\partial x_i}+\frac{\partial \sigma_{ij}}{\partial x_j}&=&0\text{,~~and}
%\label{subeq:2}
\\
    \frac{3+\delta}{2}\frac{\partial \theta}{\partial t}+\frac{\partial v_i}{\partial x_i} + \frac{\partial q_{i}}{\partial x_i}&=&0\text{.}
\end{eqnarray}
\end{subequations}
and the linearized balance equation of dynamic temperature, stress, and total heat fluxes are specified as
\begin{equation}
 \frac{3}{2}\frac{D\vartheta}{Dt}+ \frac{\delta}{3+\delta}\frac{\partial
v_{i}}{\partial x_{k}}+\frac{2\delta}{(3+\delta)(5+\delta)}\frac{\partial q_{k}}{\partial x_{k}} = \mathcal{P}^{0,0}. \label{reduced dynamic temperature 2} 
\end{equation}
\begin{equation}
    \sigma_{ij} =-2 \mu \frac{\partial v_{\langle i}}{\partial x_{j\rangle }}\text{~~~and} ~~~q_k=-\lambda \frac{\partial \theta}{\partial x_{k}}-\frac{2\lambda}{5+\delta} \frac{\partial\vartheta}{\partial x_{k}},
\end{equation}
here $\lambda=\frac{(5+\delta)^2}{25}\zeta_{11}$ is thermal conductivity of gases.
%%%%%%%%%%%%%%%%%%%%%%%%%%%%%%%%%%%%%%%%%%%%%%%%%%%%%%%
\section{\label{Linear stability analysis}Linear stability analysis}
In this section, we examine both temporal and spatial stability analyses of the two temperature models derived in section \S\,\ref{Model equation} with different coefficients given in subsections 2.3.2 and 2.3.3. Now we consider a one-dimensional process (in the x-direction) without any external forces and assume a plane wave solution of the form 
\begin{equation}
    \Phi=\Phi_a ~exp[\mathrm{i}(\omega t-kx)]
    \label{wave sol}
\end{equation}
for equations \eqref{linear equation 1} and \eqref{linear equation 2} where $\Phi=\{{\rho,v_x,\theta^{tr},\theta^{in}}\}^{T}$, $\Phi_a$ is a vector containing the complex amplitudes of the respective waves, $\omega$ and $k$ are the dimensionless frequency and the wavenumber, respectively. Substitution of the plane wave solution \eqref{wave sol} into equations \eqref{linear equation 1} and \eqref{linear equation 2} gives a system of algebraic equations $A  \Phi = 0$, where 
\begin{eqnarray}
    A(\omega,k)=\nonumber\\ 
    \begin{bmatrix}
         \mathrm{i}\omega & -\mathrm{i} k& 0 &  0 \\
         -\mathrm{i} k &\frac{4 k^2 \mu}{3}+\mathrm{i} \omega & -\mathrm{i} k &  0 \\
         0 & -\mathrm{i} k & \frac{3 \delta P^{0}_{\Pi}}{2(3+\delta)}+k^2 \zeta_{11}+\frac{3}{2}\mathrm{i} \omega &  -\frac{3 \delta P^{0}_{\Pi}}{2(3+\delta)}+k^2 \zeta_{12}   \\
         0 & 0 & -\frac{3 \delta P^{0}_{\Pi}}{2(3+\delta)}+k^2 \zeta_{21} &  \frac{3 \delta P^{0}_{\Pi}}{2(3+\delta)}+k^2 \zeta_{22}+\frac{\delta}{2}\mathrm{i} \omega
     \end{bmatrix}\nonumber\\
\end{eqnarray}
This matrix is for the model 3 coefficients; in the same way, we will be driving for the other model's coefficients.
The analogous dispersion relation found when the determinate of $ A(\omega,k)$ is zero is the relation between $\omega$ and  $k$ for non-trivial solutions. For temporal stability, a disturbance is considered in space; consequently, the wave number $k$ is assumed to be real while the frequency $\omega=\omega_{r}(k)+\mathrm{i}\omega_{i}(k)$ can be complex. The phase velocity $v_{ph}$ and damping $\alpha$
of the corresponding waves are given by
\begin{equation*}
    v_{ph}=\frac{\omega_{r}(k)}{k} ~~\text{and}~~ \alpha= \omega_i (k).
\end{equation*}
The stability of equations may be tested in two ways: temporal stability and spatial stability. Temporal stability requires damping, and thus $\omega_i (k) \geq 0$. 
If, on the other hand, $\omega_i (k) \leq 0$, a little disturbance in space will blow-up in time.\\
If a disturbance in time at a given location is considered,
the frequency $\omega$ is real, while the wave number is complex, $k = k_r (\omega) + \mathrm{i} k_i (\omega)$. Phase velocity $v_{ph}$ and damping $\alpha$ of the
corresponding waves are given by
\begin{equation*}
    v_{ph}=\frac{\omega}{k_r (\omega)}\text{,} ~~\text{and}~~ \alpha=- k_i (\omega).
\end{equation*}
For a wave traveling in positive $x$ direction $(k_r> 0)$, the
damping must be negative, while for a wave traveling in negative $x$ direction $(k_r< 0)$, the damping must be
positive $(k_i> 0)$.

\begin{figure}
\includegraphics[scale = .7]{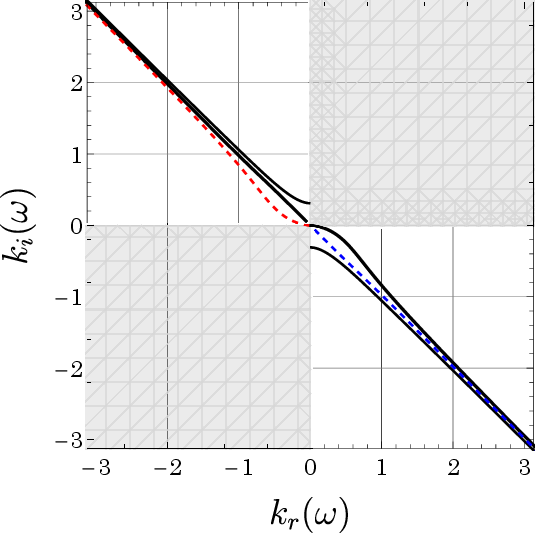}
\includegraphics[scale = .7]{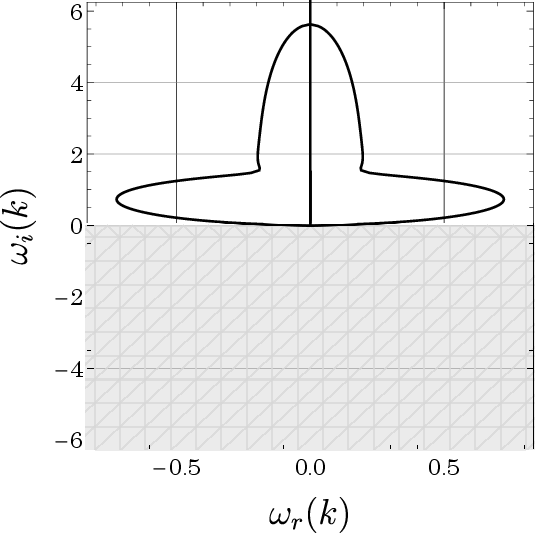}
\caption{\label{Marques_Ch4stable}Stability analysis of the two-temperature model for $CH_4$ gas with Model 1 coefficients: panel (up) spatial stability and panel (down) temporal stability.}
\end{figure}

\begin{figure}
\includegraphics[scale = .7]{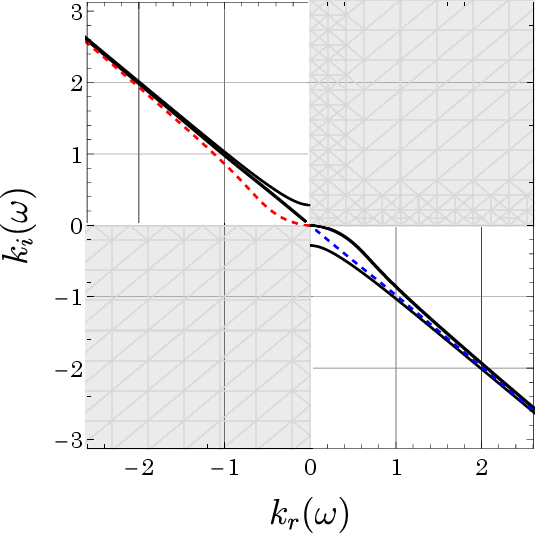}
\includegraphics[scale = .7]{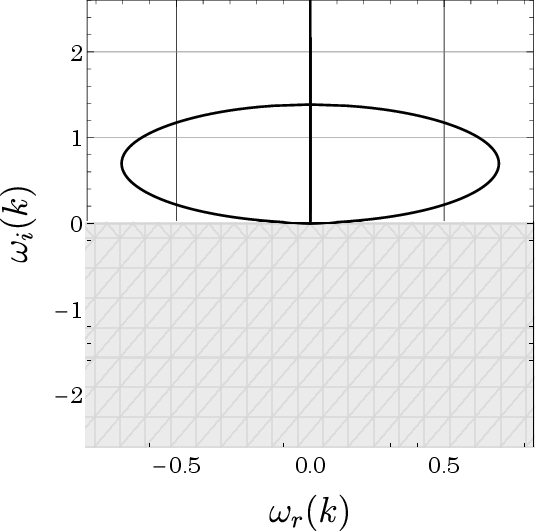}
\caption{\label{Aoki_Ch4stable}Stability analysis of the two-temperature model for $CH_4$ gas with Model 2 coefficients: panel (up) spatial stability and panel (down) temporal stability.}
\end{figure}

\begin{figure}
\includegraphics[scale = .7]{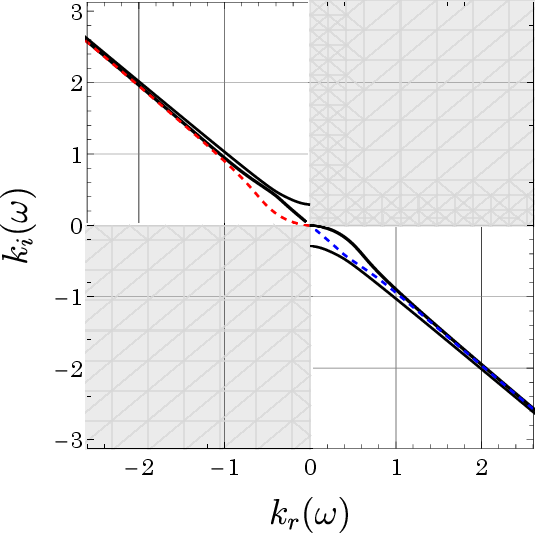}
\includegraphics[scale = .7]{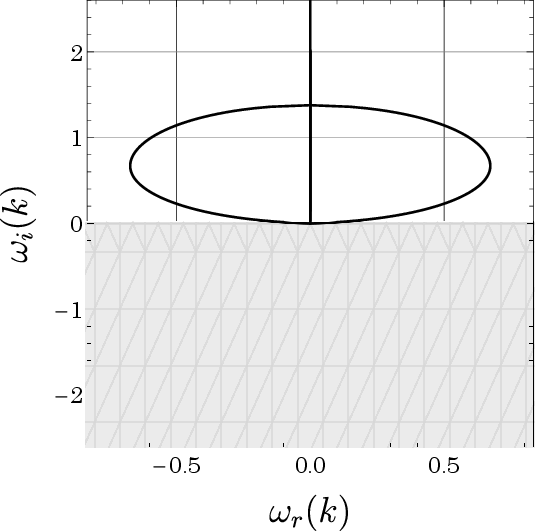}
\caption{\label{Manual_Ch4stable}Stability analysis of the two-temperature model for $CH_4$ gas with Model 3 coefficients: panel (up) spatial stability and panel (down) temporal stability.}
\end{figure}

\begin{figure}
\includegraphics[scale = .7]{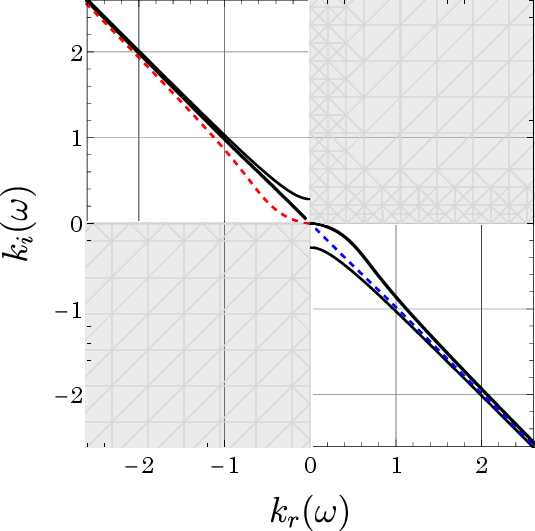}
\includegraphics[scale = .7]{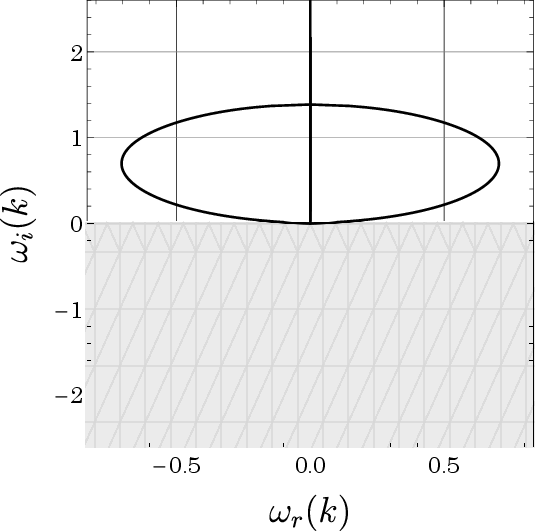}
\caption{\label{rahimi_Ch4stable}Stability analysis of the two-temperature model for $CH_4$ gas with Model 4 coefficients: panel (up) spatial stability and panel (down) temporal stability.}
\end{figure}

\begin{figure}
\includegraphics[scale = .7]{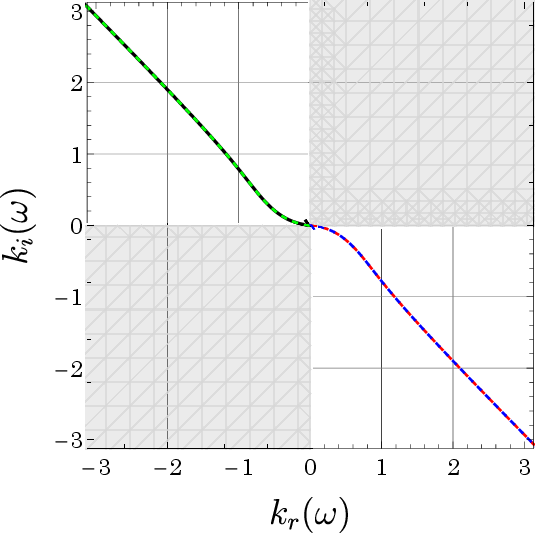}
\includegraphics[scale = .7]{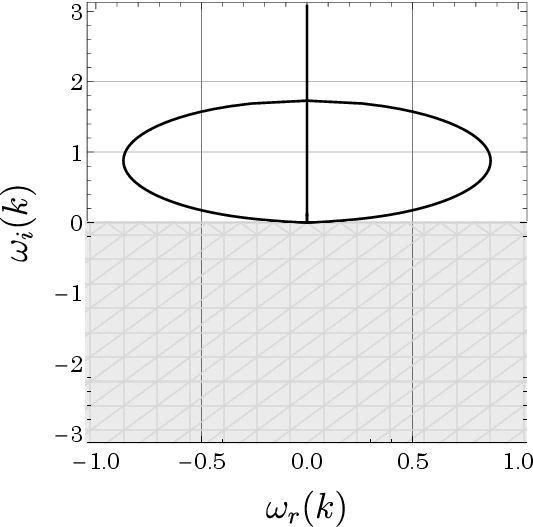}
\caption{\label{reduced_Ch4stable}Stability analysis of the two-temperature model for $CH_4$ gas with reduced model coefficients: panel (up) spatial stability and panel (down) temporal stability.}
\end{figure}

 We check how stable it is against small changes in frequency $\omega$. As we've seen, stability needs different signs of the real and imaginary parts of $k(\omega)$. So, if $k(\omega)$ is plotted in the complex plane with  $\omega$ as the parameter, the curves shouldn't touch the upper right or lower left quadrants. Figs.~\ref{Marques_Ch4stable}- \ref{reduced_Ch4stable} (up)
%Fig.~\ref{Aoki_Ch4stable}(a), Fig.~\ref{Manual_Ch4stable}(a), Fig.~\ref{rahimi_Ch4stable}(a) and Fig.~\ref{reduced_Ch4stable}(a) 
illustrate the solutions for the six-moment system, and we can see that all modes are inside the requisite stability field, indicating that none of the solutions violate the stability condition.
In  Figs.~\ref{Marques_Ch4stable}- \ref{reduced_Ch4stable} (up) we analyze the stability against a disturbance of a specific wavelength, denoted by the wave number $k$. The
figures show the damping coefficient $\alpha$ is
positive for all $k$, and it follows that the 6-moment system is stable. The two-temperature model 
equations are stable for all frequencies, as well as they are stable for disturbances of any wavelength. We can additionally show that the two-temperature model  is stable for other gases ($N_2, O_2, H_2, CO_2$) as well.
%%%%%%%%%%%%%%%%%%%%%%%%%%%%%%%%%%%%%%%%%%%%%%%%%%%%
\section{\label{Sound wave propagation}Sound wave propagation}
Wave propagation phenomena provide a valuable means to assess the validity of non-equilibrium thermodynamics theories. In this section, we focus on the study of plane harmonic waves and aim to compare the theoretical predictions of the dispersion relation with experimental data. To ensure a more manageable analysis, we restrict our investigation to the one-dimensional spatial problem.
In order to investigate the propagation in the $x$-direction of high-frequency sound waves having an angular frequency 
 and complex wavenumber $k=k_r+\mathrm{i}k_i$ 
 we write the plane wave Solution of the form equation \eqref{wave sol} and obtain a dispersion relation which is already discussed in \S\,\ref{Linear stability analysis}. The dispersion relation permits the calculation of the phase
velocity $v_{ph}$ and of the attenuation coefficient $\alpha$ in terms of the frequency $\omega$:
\begin{equation*}
    v_{ph}=\frac{\omega}{k_r (\omega)}\text{,} ~~\text{and}~~ \alpha=- k_i (\omega).
\end{equation*}
\subsection{Comparison with Experimental Data}
The dispersion relation obtained from $\det(A)=0$, in particular, the phase velocity $v_{ph}$,
attenuation factor $\alpha$ and the speed of sound as the functions of the
frequency $\omega$ are compared with the
acoustic measurements of Greenspan in nitrogen $(N_2)$ and oxygen $(O_2)$ gases \cite{greespan1959rotational}. These sound propagation measurements
were made at a temperature of $300 K$ in a $11 MHz$ double-crystal interferometer for different values of
the gas pressure.

\begin{figure}
\includegraphics[scale = .7]{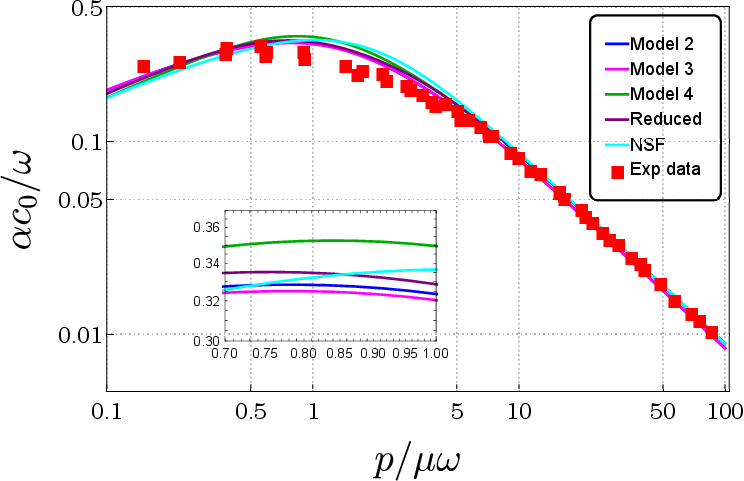}
\includegraphics[scale = .7]{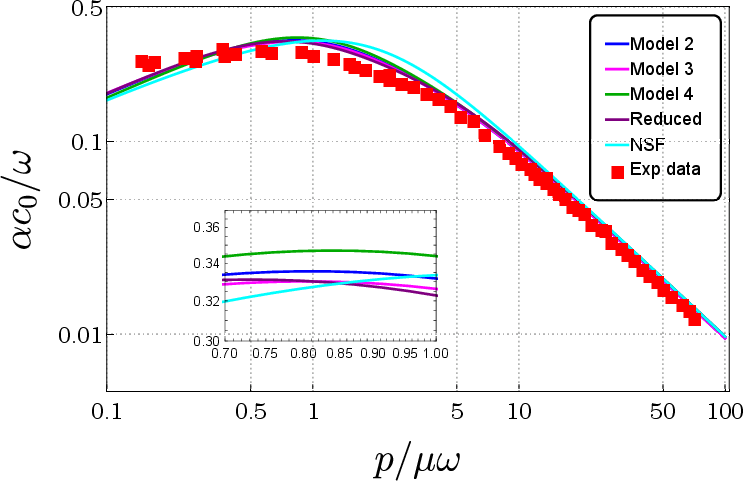}
\caption{\label{fig:attenuation}Schematics for nitrogen (up) and oxygen (down), depicting the attenuation factor $\alpha c_0/\omega$ as a function of the rarefaction parameter $p/\mu\omega$ at a temperature of $300 K$. The predictions of our theory using various coefficients and the Navier-Stokes-Fourier (NSF) theory are compared with experimental data obtained by Greenspan \cite{greespan1959rotational}.}
\end{figure}

\begin{figure}
\includegraphics[scale = .7]{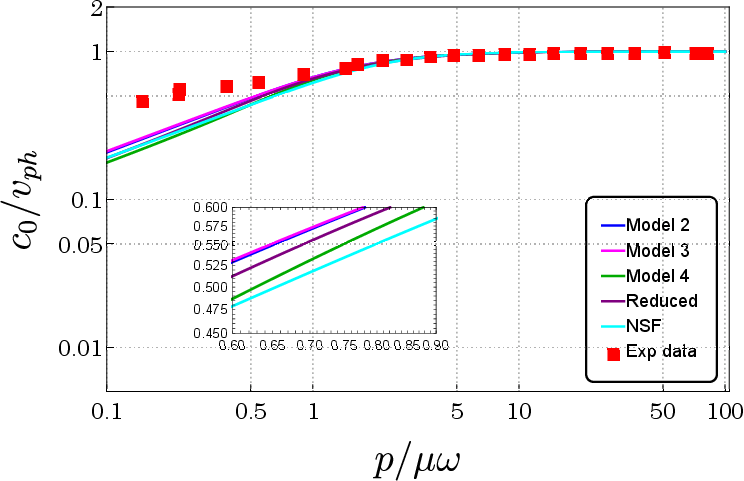}
\includegraphics[scale = .7]{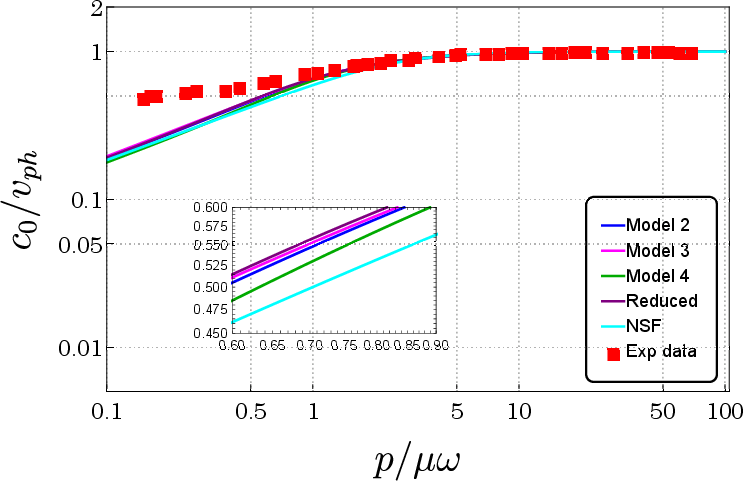}
\caption{\label{fig:phase velocity}Schematics for nitrogen (up) and oxygen (down), depicting the reciprocal speed ratio $c_0/v_{ph}$ as a function of the rarefaction parameter $p/\mu\omega$ at a temperature of $300 K$. The predictions of our theory using various coefficients and the Navier-Stokes-Fourier (NSF) theory are compared with experimental data obtained by Greenspan \cite{greespan1959rotational}.}
\end{figure}

\begin{figure}
\includegraphics[scale = .7]{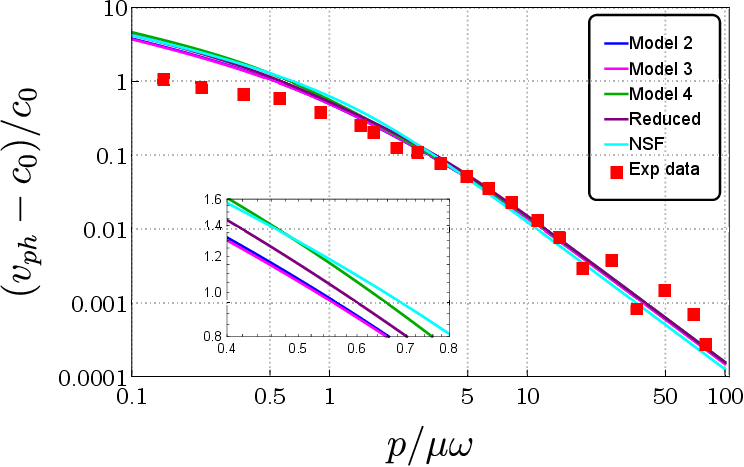}
\includegraphics[scale = .7]{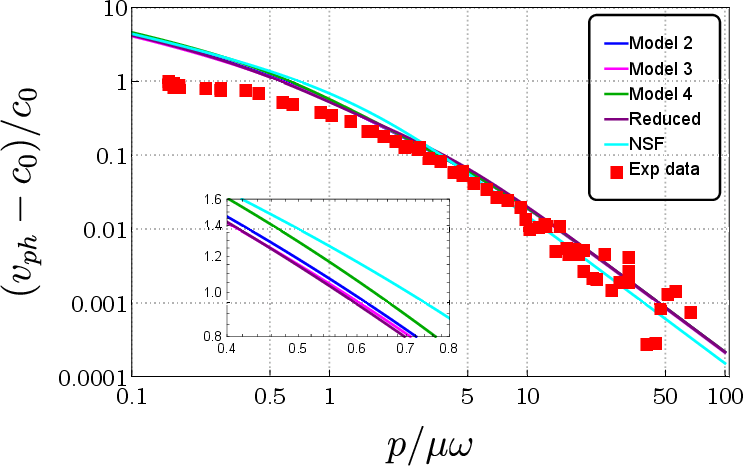}
\caption{\label{fig:speed of sound}Schematics for nitrogen (up) and oxygen (down), depicting speed of sound $(v_{ph}-c_0)/{c_0}$ as a function of the rarefaction parameter $p/\mu\omega$ at a temperature of $300 K$. The predictions of our theory using various coefficients and the Navier-Stokes-Fourier (NSF) theory are compared with experimental data obtained by Greenspan \cite{greespan1959rotational}.}
\end{figure}

In Fig.~\ref{fig:attenuation}, Fig.~\ref{fig:phase velocity} and Fig.~\ref{fig:speed of sound} the attenuation factor $\alpha c_0/\omega$, the reciprocal speed ratio  $c_0/v_{ph}$ and speed of sound $(v_{ph}-c_0)/{c_0}$ are shown
on a double logarithmic scale as a function of the non-equilibrium parameter $p/\mu\omega$, which is the ratio of collision frequency in gas and frequency of the sound, for nitrogen and
oxygen.  The solid lines represent the theoretical sound propagation results derived from
our theory and NSF (Cyan line), while the red square box is the experimental data of Greenspan
for sound waves. 
%absorption and dispersion.
From the analysis of Fig.~\ref{fig:attenuation}, Fig.~\ref{fig:phase velocity}, and Fig.~\ref{fig:speed of sound}, it can be observed that in the low-frequency range ($p/\mu\omega >>1$), both our theories and the classical Navier-Stokes-Fourier (NSF) theory exhibit excellent agreement with experimental data for sound propagation. This agreement indicates that in near-equilibrium regime, our theory and the NSF theory yield comparable results. However, as the non-equilibrium parameter decreases and we move into a transition regime ($1 < p/\mu\omega<10$), our theory shows improved performance compared to the classical NSF theory. Furthermore, in the high-frequency region ($p/\mu\omega < 1$), our theory consistently outperforms the classical NSF theory, as evident from Fig.~\ref{fig:attenuation}, Fig.~\ref{fig:phase velocity}, and Fig.~\ref{fig:speed of sound}.
%%%%%%%%%%%%%%%%%%%%%%%%%%%%%%%%%%%%%%%%%

\section{\label{Spontaneous Rayleigh-Brillouin Scattering}Spontaneous Rayleigh-Brillouin Scattering}
Spontaneous Rayleigh–Brillouin scattering (SRBs) in
gases originates from instantaneous density fluctuation.
A plane polarised light with the following parameters: intensity $I_{0}$, angular frequency $\omega_{0}$, wave vector $\mathbf{k}_{0}$, and polarisation $n_0$ impacts a fluid with the following dielectric constant $\epsilon_0$. The light's intensity, angular frequency $\omega_{s}$, wave vector $\mathbf{k}_{s}$, and polarization $n_{s}$, which was detected at distance $d$ from scattered volume a detector by the fluid's scattering at angle $\theta$ is given by

\begin{equation}
    I(\textbf{K},\omega,d)=I_{0}\frac{(\omega_{0})^4}{16 \pi^2 d^2 c^4}(n_{0}\cdot n_{s})^2 S(\textbf{K},\omega)~,\label{Intensity equation}
\end{equation}
where c is the speed of light in a vacuum, $\omega=\omega_{0}-\omega_{s}$ the shift in angular frequency and in a very close approximation, the wave vector \textbf{K} of the fluctuation being observed is the momentum transfer between the incident wave vector $\mathbf{k}_{0}$, the scattered wave vector  $\mathbf{k}_{s}$ and $S(\textbf{K},\omega)$ is dynamic structure factor. The magnitude of \textbf{K} is then approximately equal to $|\textbf{K}|=\left|\mathbf{k}_{0}-\mathbf{k}_{s}\right|=2|\mathbf{k}_{0}| sin(\theta/2)$.

The hydrodynamic equations are straightforward due to the one-dimensional, linearized, and boundary conditions-free character, and thus enable analytical solutions for the Rayleigh-Brillouin scattering problem  \cite{marques1993spectral,wu2018accuracy}. The Knudsen number ($Kn$), which is defined as the ratio of the mean free path of gas molecules to the characteristic length scale (L) of the system, known as the scattering wavelength $2\pi/\textbf{K}$, is used to describe the spectrum of scattered light.
The linearized and one-dimensional form of our system is:

\begin{subequations}
\label{linear_equation_11}
\begin{eqnarray}
   \frac{\partial \hat{\rho}}{\partial \hat{t}}+\frac{\partial \hat{v}}{\partial \hat{x}}&=&\delta(\hat{t})\text{,}
 %\label{subeq:1}
\\
  \frac{\partial \hat{v}}{\partial \hat{t}}+\frac{\partial \hat{\rho}}{\partial\hat{x}}+\frac{\partial \hat{\theta}^{tr}}{\partial \hat{x}}+\frac{\partial \hat{\sigma}}{\partial \hat{x}}&=&0\text{,}
%\label{subeq:2}
% %\\
%     \frac{(3+\delta)}{2}\frac{\partial \hat{\theta}}{\partial \hat{t}}+\frac{\partial \hat{v}}{\partial \hat{x}}+\frac{\partial \hat{q}}{\partial \hat{x}}&=&0\text{,}
 \\
\frac{3}{2} \frac{\partial \hat{\theta}^{tr}}{\partial \hat{t}}+\frac{\partial \hat{v}}{\partial \hat{x}}+\frac{\partial \hat{q}^{tr}}{\partial \hat{x}}&=&\mathcal{P}^{0,0}\text{,}
\\
\frac{\delta}{2} \frac{\partial \hat{\theta}^{in}}{\partial\hat{t}}+ \frac{\partial \hat{q}^{in}}{\partial \hat{x}}&=&\mathcal{P}^{0,1}\text{,}
\end{eqnarray}
\end{subequations}
and constitutive equations
\begin{subequations}
\label{linear_equation_22}
\begin{eqnarray}
   \hat{\sigma} &=&-\frac{4}{3} Kn \frac{\partial \hat{v}}{\partial \hat{x}}\text{,}
 %\label{subeq:1}
\\
 \hat{q}^{in}&=&-\zeta_{12} \frac{\partial\hat{\theta}^{tr}}{\partial \hat{x}}-\zeta_{22} \frac{\partial\hat{\theta}^{in}}{\partial \hat{x}}\text{,~~and}
%\label{subeq:2}
\\
    \hat{q}^{tr}&=&-\zeta_{11} \frac{\partial\hat{\theta}^{tr}}{\partial \hat{x}}-\zeta_{12} \frac{\partial\hat{\theta}^{in}}{\partial \hat{x}}\text{.}
\end{eqnarray}
\end{subequations}
Equations \eqref{linear_equation_11} and \eqref{linear_equation_22} are transferred into the following matrix form by using the Laplace transform for the temporal variable $t$ and the Fourier transform for the spatial variable $x$ in the spontaneous SRB \cite{marques1993spectral}:
\begin{widetext}
\begin{equation}
 \begin{bmatrix}
         -\mathrm{i} \omega & 2 \pi \mathrm{i} & 0   & 0\\
          2 \pi \mathrm{i} & -\left(\mathrm{i} \omega-\frac{8 \pi \mathrm{i}}{3}Kn\right) &  2 \pi \mathrm{i}  & 0\\
        % 0 & 2 \pi \mathrm{i} & -\frac{(3+\delta)}{2}\mathrm{i} \omega &  4\pi^2 (\zeta_{11}+\zeta_{12})   & 4\pi^2 (\zeta_{12}+\zeta_{22})\\
         0 & 2 \pi \mathrm{i} &  -\frac{3}{2}\mathrm{i} \omega+4\pi^2 \zeta_{11}  & 4\pi^2 \zeta_{12}\\
         0 & 0 &   4\pi^2 \zeta_{12}  & -\frac{\delta}{3}\mathrm{i} \omega+4\pi^2 \zeta_{22}\\
     \end{bmatrix}
     \begin{bmatrix}
         \hat{\rho}\\ 
         \hat{v} \\
        % \hat{\theta} \\
         \hat{\theta}^{tr} \\
         \hat{\theta}^{in} \\
     \end{bmatrix}=
     \begin{bmatrix}
         1 \\ 
         0  \\
       % 0  \\
         \mathcal{P}^{0,0}  \\
         \mathcal{P}^{0,1}  \\
     \end{bmatrix}, \label{non-homogeneous matrix equations}
\end{equation}
\end{widetext}
where $\omega$ is the angular frequency. 
% In the spontaneous Rayleigh–Brillouin scattering, we have $1$ due to the Laplace transform of the initial
% density perturbation. 
For the spectrum of the density fluctuations, $\hat{\rho}$, we solve the non-homogeneous matrix equations \eqref{non-homogeneous matrix equations} and obtain the spontaneous Rayleigh–Brillouin scattering.
%%%%%%%%%%%%%%%%%%%%%%%%%%%%%%%%%%%%%
\subsection{Results}
%\subsection{Comparison with experimental data for $CH_4$}
In this section, analytic results for
the light scattering spectrum are derived for methane ($CH_4$) and compared with the predictions of
the extended hydrodynamic theory of Hammond and Wiggins \cite{hammond1976rayleigh}, which are entirely consistent with empirical data. To apply the two-temperature model to the results obtained in the previous section, certain conditions must be satisfied by the polyatomic gas. These conditions include spherical symmetry of the gas molecules and a specific heat capacity ratio close to $4/3$. Methane at ordinary temperatures approximately meets these requirements.
By solving the non-homogeneous matrix equations \eqref{non-homogeneous matrix equations} for the spectrum of the density fluctuations
$\hat{\rho}$, we obtain the SRB spectra.
An SRB spectrum typically consists of a central Rayleigh peak and two Brillouin side peaks at 
equidistant from the central Rayleigh peak. These Brillouin side peaks are located at $\sqrt{\frac{\gamma}{2}}$, where $\gamma$ is the
ratio of heat capacity. In the typical spectra of the SRBs, one can identify the contributions from the
central Rayleigh peak and the Brillouin side peaks. 

We show the scattered light spectrum from $CH_4$ gas ($293 K$ and 1.013 bar) for different values of $y$, which is a measure of the ratio of the wavelength of the observed fluctuation with the collision mean free path
\begin{equation}
  y=\frac{1}{\sqrt{2}}\left(\frac{K T_0}{m}\right)^{1/2} \frac{\rho_{0}}{\mu_{0} K} ,  
\end{equation}
with the reduced frequency $x$ is given by
\begin{equation}
  x=\sqrt{\frac{2}{3}} \frac{\omega}{\mu_{0} K}.  
\end{equation}

\begin{figure}
\includegraphics[scale = .7]{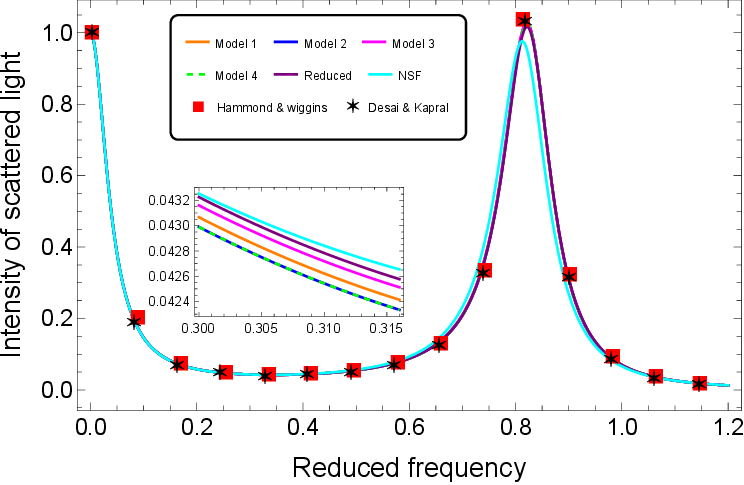}
\caption{\label{Figure01}The scattered light spectrum from $CH_4$ for $y=18.27$ and spectrum has been normalized at the Rayleigh peak.}
\end{figure}

\begin{figure}
\includegraphics[scale = .7]{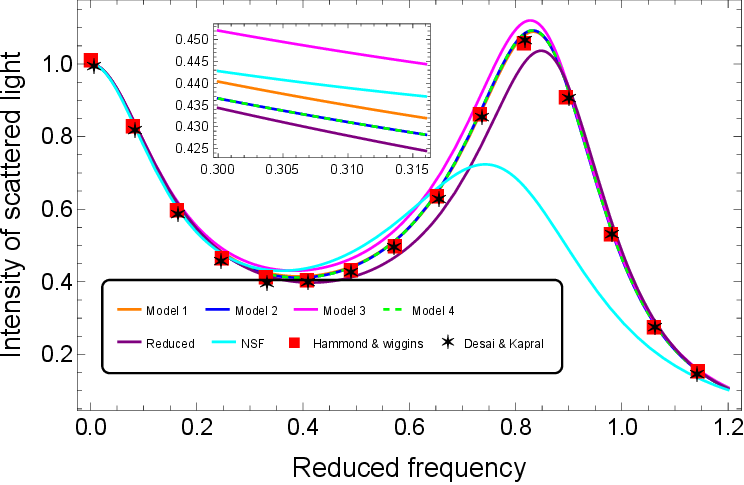}
\caption{\label{Figure02}The scattered light spectrum from $CH_4$ for $y=4.46$ and spectrum has been normalized at the Rayleigh peak.}
\end{figure}

\begin{figure}
\includegraphics[scale = .7]{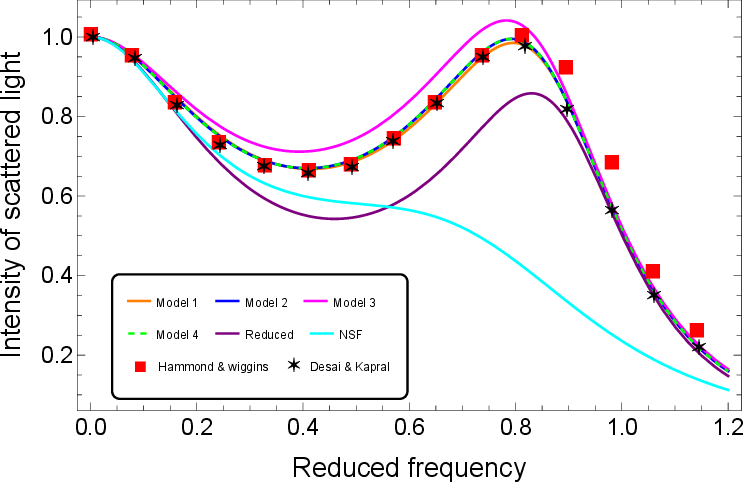}
\caption{\label{Figure03}The scattered light spectrum from $CH_4$ for $y=2.70$ and spectrum has been normalized at the Rayleigh peak.}
\end{figure}

In Figs.~\ref{Figure01}-\ref{Figure03}, we compare the light scattering results obtained from our two-temperature model equation with the predictions of the extended hydrodynamic theory proposed by Hammond and Wiggins \cite{hammond1976rayleigh}. The comparisons are made for three different values of parameter ``$y$," specifically $18.27$, $4.46$, and $2.70$. Figures display the spectral predictions of the two-temperature model represented by solid and dashed lines, while the solid line (cyan color) represents the NSF model. Additionally, the red square symbols represent the spectral prediction of the extended hydrodynamic theory, which, according to Hammond and Wiggins, exhibits excellent agreement with the experimental data (the curves are normalized at the zero frequency shift) and the diamond represent the 
prediction of the translational hydrodynamic theory from Desai and Kapral \cite{desai1972translational}.

From Figs.~\ref{Figure01}-\ref{Figure03}, we can conclude that our two-temperature model equation accurately describes the Rayleigh-Brillouin spectrum in polyatomic ideal gases with rotational energy. Moreover, it demonstrates a smooth and precise transition between the hydrodynamic and kinetic regimes. This ability to provide a seamless connection between these two regimes enhances the model's capability to capture the behavior of the gas across a wide range of conditions.

Upon analyzing the results presented in Figures~\ref{Figure01}-\ref{Figure03}, we observe that the reduced model provides favorable outcomes for $y = 18.27$ and $y = 4.46$, but exhibits slight discrepancies for $y = 2.70$. In contrast, the NSF approach fails to accurately predict the scattered light's spectrum from methane for $y = 4.46$ and $y = 2.70$. Based on these comparisons, we can conclude that the two-temperature model proposed in this work is well-suited for analyzing light scattering experiments from methane.

It is important to note that the two-temperature model equation, as presented in this study, does not account for the vibrational degrees of freedom of the molecules, unlike the extended hydrodynamic model of Hammond and Wiggins \cite{hammond1976rayleigh}. Despite this limitation, the two-temperature model shows promise in providing valuable insights and predictions for light scattering phenomena in polyatomic ideal gases with rotational energy, as demonstrated by its agreement with the extended hydrodynamic theory in certain cases.
%%%%%%%%%%%%%%%%%%%%%%%%
%%%%%%%%%%%%%%%%%%%%%%%%%%%%%%%%%%
\section{\label{Wall Boundary Conditions} Wall Boundary Conditions}
The incorporation of the second law of thermodynamics is critical in determining appropriate wall boundary conditions for gas dynamics models. It is imperative that the boundary conditions satisfy the second law to ensure that the entropy generation at the interface is positive. \\
Our approach to derive the boundary conditions is to evaluate the entropy generation at the interface and formulate the boundary conditions as phenomenological laws. The entropy generation can be computed using the entropy balance equation and integrating directly over the interface while applying Gauss's theorem.\\
The entropy production rate at the boundary $\Sigma_{\mathrm{w}}$ is given by the difference between the entropy
fluxes into and out of the surface \cite{de2013non}, i.e.,
\begin{equation}
    \Sigma_{\mathrm{w}} = \left(h_{k}-\frac{q^{w}_{k}}{\theta^{w}}\right)n_k .\label{wall entropy production}
\end{equation}
Here, $n_k$ is the unit normal pointing from the boundary into the gas, $q^{w}_{k}$
 denotes the heat flux in the
wall at the interface, and $\theta^w$ denotes the temperature of the wall at the interface. Here, the wall is
assumed to be a rigid Fourier heat conductor, with the entropy flux $\frac{q^{w}_{k}}{\theta^{w}}$.\\
At the interface, the total fluxes of mass, momentum, and energy are continuous, due to the conservation of these quantities,

\begin{subequations}
\label{flux on interface}
\begin{eqnarray}
   v_{k}n_{k}= v^{w}_{k}n_{k} &=&0\text{,}
\label{mass  flux}
\\
 ((p+\Pi) \delta _{ik}+\sigma_{ik})n_{k}&=&p^{w}n_{i}\text{,~~and}
\label{momentum flux}
\\
   ((p+\Pi)v_{k} +\sigma_{ik} v_{i}+q_{k})n_{k} &=&(p^{w}v^{w}_{k}+q^{w}_{k}) n_{k}\text{.}
   \label{energy flux}
\end{eqnarray}
\end{subequations}
where all quantities with superscript $w$ refer to wall properties, and the others refer to the gas
properties. To proceed, we combine entropy generation and continuity conditions by eliminating
the heat flux in the wall $q^{w}_{k}$
 and the pressure $p^{w}$, and find, after insertion of the entropy flux (\ref{entropy flux}),
\begin{multline}
     \Sigma_{\mathrm{w}}=  -\left[\frac{q_{k}}{\theta\theta^w}\mathcal{T}+\frac{2}{5+\delta}\frac{\vartheta}{\theta(\theta+\vartheta)}q_{k}\right.\\ \left.
     +\frac{3+\delta}{5+\delta}\frac{\vartheta}{\theta(\theta+\vartheta)}Q_k+\sigma_{ik}\frac{\mathcal{V}_i}{\theta^{w}}\right]n_k.
    \label{entropy flux 1}
\end{multline}
Here, $\mathcal{V}_{i}=v_{i}-v^{w}_{i} $ is the slip velocity, with $\mathcal{V}_{i} n_i=0$, and $\mathcal{T}=\theta-\theta^w$ is the temperature jump.
To write the entropy generation properly as the sum of products of forces and fluxes, it is necessary
to decompose the stress tensor and heat flux into their components in the normal and tangential directions as \cite{rana2016thermodynamically}
\begin{subequations}
\label{tensor eq1}
\begin{eqnarray}
   q_{k} &=&q_{n}n_{k}+\Bar{q}_{k}\text{,}
%\label{mass  flux}
\\
  Q_{k} &=&Q_{n}n_{k}+\Bar{Q}_{k}\text{,~~and}
%\label{momentum flux}
\\
   \sigma_{ik} &=&\sigma_{nn}\left(\frac{3}{2}n_{i}n_{k}-\frac{1}{2}\delta_{ik}\right)+\Bar{\sigma}_{ni}n_{k}+\Bar{\sigma}_{nk}n_{i}+\tilde{\sigma}_{ik}\text{.}
  % \label{energy flux}
\end{eqnarray}
\end{subequations}
where $q_n=q_{i}n_{i}, \sigma_{nn}=\sigma_{jl}n_{j}n_{l}$, and
\begin{subequations}
\label{tensor eq2}
\begin{eqnarray}
  \Bar{q}_{k} &=&q_{k}-q_{n}n_{i}\text{,}
\\
  \Bar{Q}_{k} &=&Q_{k}-Q_{n}n_{i}\text{,}
\\
    \Bar{\sigma}_{ni}&=&\sigma_{il}n_{l}-\sigma_{nn}n_{i}\text{,~~and}
\\
\tilde{\sigma}_{ik} &=&\sigma_{ik}-\sigma_{nn}\left(\frac{3}{2}n_{i}n_{k}-\frac{1}{2}\delta_{ik}\right)-\Bar{\sigma}_{ni}n_{k}-\Bar{\sigma}_{nk}n_{i} \text{,}
\end{eqnarray}
\end{subequations}

such that $\Bar{q}_{l}n_l=\Bar{\sigma}_{nl}n_l=\tilde{\sigma}_{ll}=\tilde{\sigma}_{ik}n_k=\tilde{\sigma}_{ik}n_i=0$.
Substituting equations (\ref{tensor eq1}) and (\ref{tensor eq2}) into equation (\ref{entropy flux 1}), the entropy generation can be written
as a sum of three contributions:

\begin{multline}
    \Sigma_{\mathrm{w}}=-\frac{\mathcal{T}}{\theta \theta ^{w}}\left[ q_{n}-\frac{2}{5+\delta }
\frac{\vartheta }{(\theta +\vartheta )}q_{n}-\frac{3+\delta }{5+\delta }
\frac{\vartheta }{(\theta +\vartheta )}Q_{n} \right.\\ \left.
+\Bar{\sigma}_{ni}\mathcal{V}_{i}
\right] 
-\frac{1}{\theta \theta ^{w}}\frac{\vartheta \theta }{(\theta +\vartheta )}
\left( \frac{2}{5+\delta }q_{n}+\frac{3+\delta }{5+\delta }Q_{n}\right) -
\frac{\mathcal{V}_{i}}{\theta }\Bar{\sigma}_{ni}.
\label{entropy flux 2}
\end{multline}
For a positive entropy production, we find the phenomenological boundary conditions for the two-temperature model as 
\begin{widetext}
\begin{subequations}
\label{PBC}
\begin{eqnarray}
q_{n}-\frac{2}{5+\delta }\frac{\vartheta }{\theta +\vartheta }q_{n}-\frac{%
3+\delta }{5+\delta }\frac{\vartheta }{(\theta +\vartheta )}Q_{n}+\Bar{\sigma%
}_{ni}\mathcal{V}_{i} &=&- \eta_{11}\mathcal{T}-\eta_{12}\frac{\vartheta }{(\theta +\vartheta )}\theta\text{,}\\
\frac{2}{5+\delta }q_{n}+\frac{3+\delta }{5+\delta }Q_{n} &=&- \eta_{12}\mathcal{T}-\eta_{22}\frac{\vartheta }{(\theta +\vartheta )}\theta\text{,}\\
\Bar{\sigma}_{ni}&=&-\Xi\mathcal{V}_i
\text{.} 
\end{eqnarray}
\end{subequations}
\end{widetext}
Here, the matrices
\begin{equation}
    \eta=\begin{bmatrix}
        \eta_{11} & \eta_{12}\\ 
         \eta_{12} & \eta_{22} 
     \end{bmatrix}
\end{equation}
 is  a symmetric non-negative definite matrix of  Onsager resistivity coefficients and $\Xi\geq0$, which can be obtained either from experiments
or from kinetic theory models, as we shall show in next sections.
\subsection{Comparison with Rahimi and Struchtrup \cite{rahimi2016macroscopic}}
The Onsager resistivity coefficients $(\eta_{ij})$ appearing in the boundary conditions Eqs.~(\ref{PBC}) are obtained from kinetic theory in the asymptotic limit of small dynamic temperature $(\vartheta \to 0)$.

\begin{subequations}
\label{PBC coeffi bc}
\begin{eqnarray}
    \eta_{11}=&&\frac{
\chi }{2-\chi }\frac{(4+\delta )}{2}p\sqrt{\frac{2}{\pi \theta }}+O(\min (\epsilon ^{\alpha },\epsilon ))\label{eta_{11}} \text{,}
\\
    \eta_{12}=&&\frac{1}{2}\frac{\chi }{2-\chi }p\sqrt{\frac{2}{\pi
\theta }}+O(\min (\epsilon ^{\alpha },\epsilon ))\label{eta_{12}}\text{,}
\\
   \eta_{22}=&& \frac{\chi }{2-\chi }\sqrt{\frac{2}{\pi \theta }}p \left[\frac{15+4\delta}{2\delta}\frac{(3+\delta)}{(5+\delta)}+\frac{1}{(3+\delta) }\right]\nonumber\\&&
   +O(\min (\epsilon ^{\alpha },\epsilon ))\label{eta_{22}} \text{,}\\ 
   \Xi=&&\frac{\chi }{2-\chi }\sqrt{\frac{2}{\pi \theta }}p+O(\min (\epsilon ^{\alpha },\epsilon ))\text{,}
\end{eqnarray}
\end{subequations}
where $\chi$ is the wall accommodation coefficients, specifying the level of
accommodation of the particle on the wall. Full accommodation is specified by 
$\chi=1$ and the pure specularly reflected particles are described by $\chi=0$. Solving boundary conditions Eqs.~(\ref{PBC}) for $q_{n}$ and $Q_{n}$ and after that linearized and we obtain 

\begin{eqnarray}
    q_{n}=&&- \eta_{11} \mathcal{T-}\eta_{12}\vartheta,
\label{BC heat linear}
\\
% \end{eqnarray}
% \begin{eqnarray}
    Q_{n} =&&-\frac{5+\delta}{3+\delta}\left[ \eta_{12}-\frac{2}{5+\delta } \eta_{11}\right]\mathcal{T}\nonumber\\&&
-\frac{5+\delta}{3+\delta}\left[\eta_{22}-\frac{2}{5+\delta } \eta_{12}\right] \vartheta.
\label{BC heatdiff linear}
\end{eqnarray}
Now substitute the value of coefficients  from equation \eqref{PBC coeffi bc} in phenomenological boundary conditions \eqref{PBC} and linearized, dimensionless, we get the following boundary conditions
\begin{subequations}
\label{PBC linear}
\begin{eqnarray}
\hat{q}_{n}=&&-\frac{
\chi }{2-\chi }\sqrt{\frac{2}{\pi}} \frac{(4+\delta )}{2} \mathcal{\hat{T}}
-\frac{\chi }{2-\chi }\sqrt{\frac{2}{\pi
 }}\frac{1}{2}\hat{\vartheta}\text{,}\\
\hat{Q}_{n} =&&\frac{1}{2}\frac{\chi }{2-\chi }\sqrt{\frac{2}{\pi
 }}\mathcal{\hat{T}}
-\frac{\chi }{2-\chi }\sqrt{\frac{2}{\pi }}\left[ \frac{15+4\delta}{2\delta}+\frac{2}{(3+\delta)^2 }\right] \hat{\vartheta}\text{,}\nonumber\\
\\
\hat{\Bar{\sigma}}_{ni}=&&-\frac{\chi }{2-\chi }\sqrt{\frac{2}{\pi }}\hat{\mathcal{V}}_i.
\end{eqnarray}
\end{subequations}

%%%%%%%%%%%%%%%%%%%%%%%%%%%%%%%%%%%%%%%%%%%%%%%%%%%
\subsection{Reduced model boundary condition}
For Reduced model boundary condition $Q_{n}=0$ i.e., 
\begin{equation}
    \eta_{12}=\frac{2}{5+\delta}\eta_{11}~~~~~~~ \text{and}~~~~~~ \eta_{22}=\frac{4}{(5+\delta)^2}\eta_{11}.
    \label{reduced bc  coefficients}
\end{equation}
Upon substituting the values of these coefficients from equation \eqref{reduced bc coefficients} into the phenomenological boundary conditions \eqref{PBC}, and linearizing and dimensionless the equations, we obtain the following boundary conditions
\begin{subequations}
\label{BC heat and stress linear}
\begin{eqnarray}
    \hat{q}_{n}=&&- \frac{
\chi }{2-\chi }\sqrt{\frac{2}{\pi}} \frac{4+\delta}{2} \mathcal{\hat{T}} -\frac{
\chi }{2-\chi }\sqrt{\frac{2}{\pi}} \frac{4+\delta }{5+\delta} \hat{\vartheta} \text{,}\\
    \hat{\Bar{\sigma}}_{ni}=&&-\frac{\chi }{2-\chi }\sqrt{\frac{2}{\pi }}\hat{\mathcal{V}}_i.
\end{eqnarray}
\end{subequations}
%%%%%%%%%%%%%%%%%%%%%%%%%%%%%%%%%%%%%%%%%%%%%%%%%%%%%%%%%
\section{\label{Flow between two parallel plates}Heat transfer between two parallel plates}
This section focuses on the validation of our wall boundary conditions through an investigation of a fundamental problem involving steady-state conductive heat transfer through stationary rarefied gases confined between parallel plates. To achieve this, we employ the two-temperature model within the framework of rarefied gas dynamics.
\subsection{Problem settings and reduced moment equations}
\begin{figure}
\includegraphics[scale = .8]{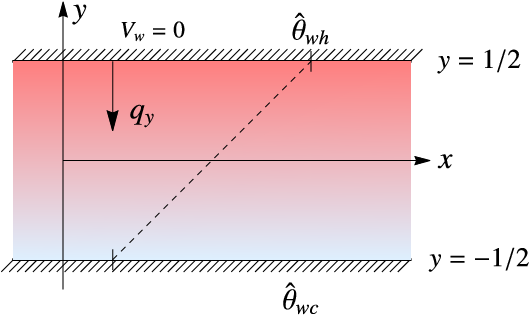}
\caption{\label{schematic_heat}The schematic illustrates heat conduction, featuring a channel with two stationary parallel walls having distinct temperatures.}
\end{figure}

In this study, we investigate a steady-state gas flow between two parallel plates that are infinitely long and fixed perpendicular to the $y$-axis at $y = \pm 1/2$ (see Fig.~\ref{schematic_heat}). The temperature of the walls differs, and the flow properties and variables solely depend on the y-direction. The fluid is assumed to be stationary in this case, and any flow caused by density changes in an unsteady state is ignored.
The two temperature model equation \eqref{linear equation 1} can be derived for one-dimensional channel flows by omitting partial derivatives with respect to $x$ and $z$. 
Thus, the conservation laws are modified in this form
\begin{equation}
\frac{\partial \hat{\sigma}_{xy}}{\partial \hat{y}}=0\text{, \quad } \\
   \frac{\partial \hat{\rho}}{\partial\hat{y}}+\frac{\partial \hat{\theta}}{\partial \hat{y}}+\frac{\partial \hat{\vartheta}}{\partial \hat{y}}+\frac{\partial \hat{\sigma}_{yy}}{\partial \hat{y}}=0,~~
    \text{and\quad } \\ 
 \frac{\partial \hat{q}_{y}}{\partial \hat{y}}=0,
   \label{linear equation and steady}
\end{equation}
and equations of dynamic temperature
\begin{equation}
 \left(\frac{2\delta}{(3+\delta)(5+\delta)}\right)\frac{\partial \hat{q}_{y}}{\partial y}+\left(\frac{\delta}{5+\delta}\right)\frac{\partial \hat{Q}_{y}}{\partial y} =
\mathcal{P}^{0,0} . \label{dynamic temperature and steady} 
\end{equation}
The equation of stress tensor $\hat{\sigma}_{xy}$, total heat flux $\hat{q}_{y}$ and heat difference $Q_{y}$ are
\begin{subequations}
\label{linear1D steady}
\begin{eqnarray}
\hat{\sigma}_{xy}&=&-\frac{\mathrm{4 Kn}}{3} \frac{\partial \hat{v}}{\partial \hat{y}}\text{,} 
\\
\hat{q}_{y}&=&-(\zeta_{11}+2\zeta_{12}+\zeta_{22}) \frac{\partial\hat{\theta}}{\partial \hat{y}}\nonumber\\&&
-\left[(\zeta_{11}+\zeta_{12})-\frac{3}{\delta}(\zeta_{12}+\zeta_{22})\right]\frac{\partial\hat{\vartheta}}{\partial \hat{y}} \text{,}
\\
\hat{Q}_{y}&=&-\left[(\zeta_{11}-\frac{5}{\delta}\zeta_{12})+(\zeta_{12}-\frac{5}{\delta}\zeta_{22})\right]\frac{\partial\hat{\theta}}{\partial \hat{y}}\nonumber\\&&
-\left[(\zeta_{11}-\frac{5}{\delta}\zeta_{12})-\frac{3}{\delta}(\zeta_{12}-\frac{5}{\delta}\zeta_{22})\right]\frac{\partial\hat{\vartheta}}{\partial \hat{y}}\text{.\qquad }
\end{eqnarray}
\end{subequations}
The linear system can be easily solved analytically, yielding the general solution that involves seven constants to be determined. One of the constants depends on the average density between the plates: upon setting the range of $y$ to be $[-1/2, 1/2]$
 (so that $L = 1$), we assign the average density as
\begin{equation}
     \int_{-1/2}^{1/2} \hat{\rho}(y) \,dy =0.
\end{equation}
The remaining six constants can be determined through six boundary conditions. In the one-dimensional setting, each boundary contributes three boundary conditions, as specified by equation \eqref{PBC linear}. Here, we present the boundary conditions specifically for the upper wall.
\begin{subequations}
\label{PBC heat linear BC}
\begin{eqnarray}
\hat{q}_{y}=&&-\frac{
\chi }{2-\chi }\sqrt{\frac{2}{\pi}} \frac{(4+\delta )}{2} \mathcal{\hat{T}}
-\frac{\chi }{2-\chi }\sqrt{\frac{2}{\pi
 }}\frac{1}{2}\hat{\vartheta}\text{,}\\
\hat{Q}_{y}=&&\frac{1}{2}\frac{\chi }{2-\chi }\sqrt{\frac{2}{\pi
 }}\mathcal{\hat{T}}
-\frac{\chi }{2-\chi }\sqrt{\frac{2}{\pi }}\left[ \frac{15+4\delta}{2\delta}+\frac{2}{(3+\delta)^2 }\right] \hat{\vartheta}\text{,}\nonumber\\
\\
\hat{\Bar{\sigma}}_{xy}=&&-\frac{\chi }{2-\chi }\sqrt{\frac{2}{\pi }}\hat{\mathcal{V}}_2.
\end{eqnarray}
\end{subequations}

To apply the boundary conditions on the lower solid wall, it is sufficient to make the following parameter replacement:
\begin{equation}
    \hat{q}_{y} \to-\hat{q}_{y}\text{,\qquad }\\
    \hat{Q}_{y} \to-\hat{Q}_{y}\text{,\qquad }\\
    \hat{\Bar{\sigma}}_{xy} \to-\hat{\Bar{\sigma}}_{xy}\text{.\qquad }
\end{equation}
%%%%%%%%%%%%%%%%%%%%
\subsection{Results}
\begin{figure}
\includegraphics[scale =0.6]{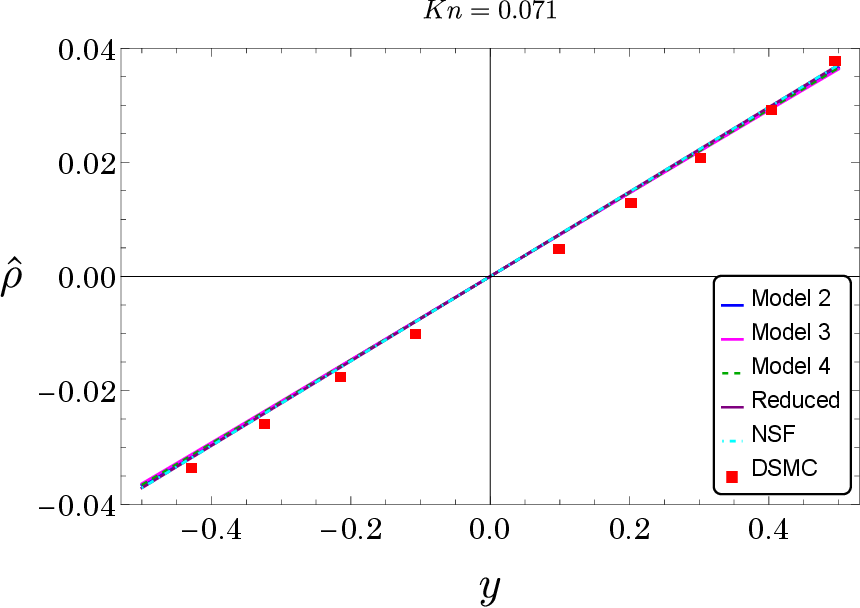}
\includegraphics[scale =0.6]{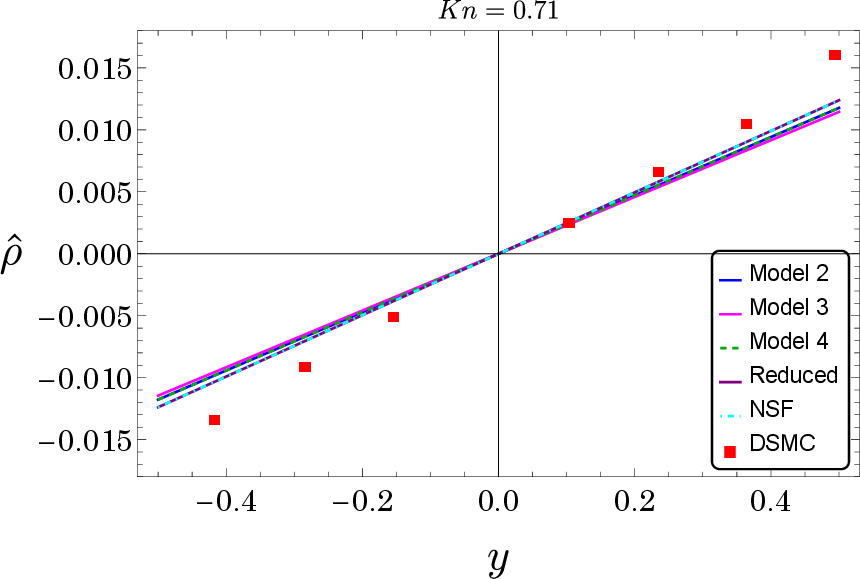}
\caption{\label{density}Comparison of density profiles for $Kn$ numbers equal to $0.071$ (up) and $0.71$ (down).  Results shown are obtained from: our theory (solid, dashed line); NSF equations (dot dashed cyan line); DSMC method (red square box)\cite{tantos2015conductive}.}
\end{figure}

\begin{figure}
\includegraphics[scale =.6]{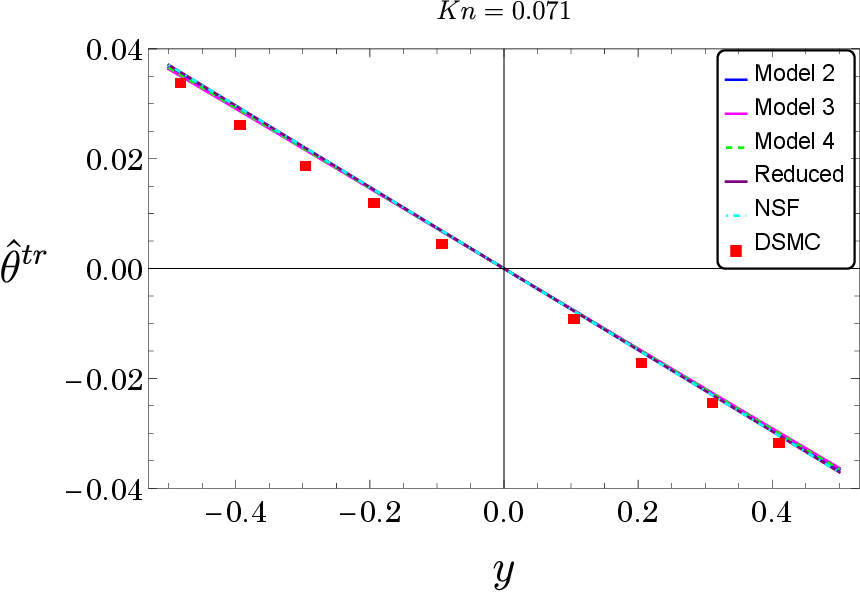}
\includegraphics[scale =.6]{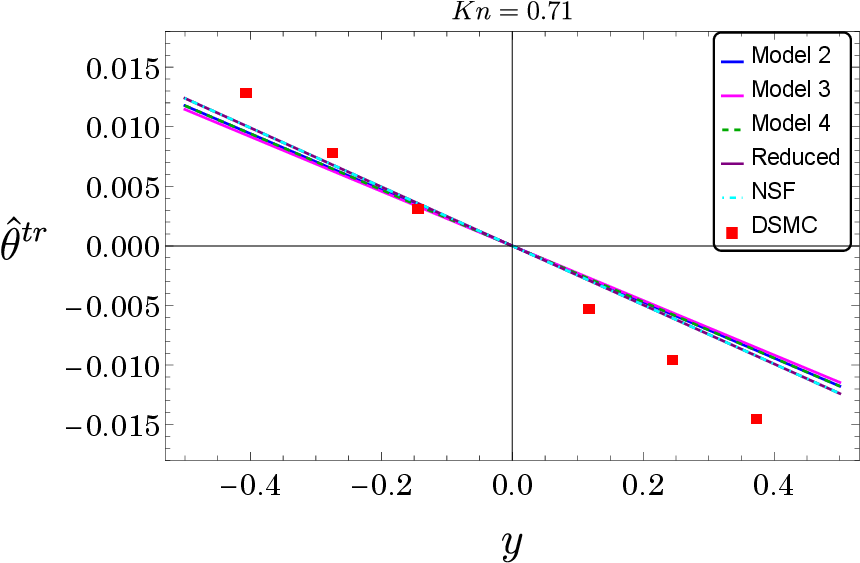}
\caption{\label{thetatr}Comparison of translational temperature for $Kn$ numbers equal to $0.071$ (up) and $0.71$ (down).  Results shown are obtained from: our theory (solid, dashed line); NSF equations (dot dashed cyan line); DSMC method (red square box)\cite{tantos2015conductive}.}
\end{figure}

\begin{figure}
\includegraphics[scale = .6]{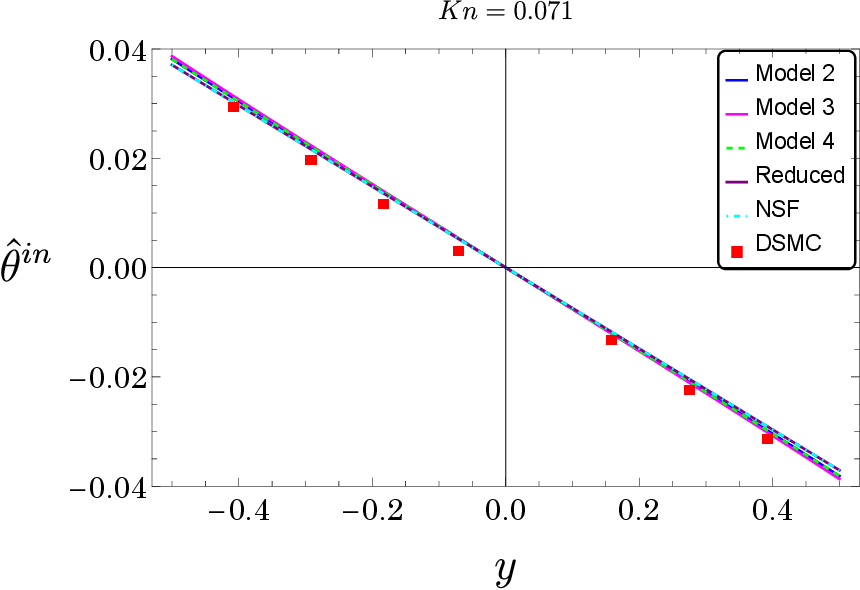}
\includegraphics[scale = .6]{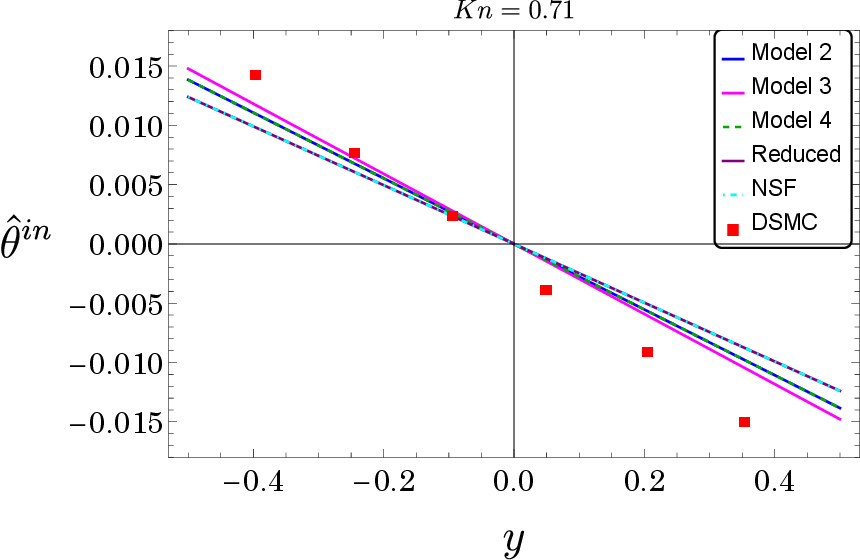}
\caption{\label{thetain}Comparison of internal temperature for $Kn$ numbers equal to $0.071$ (up) and $0.71$ (down).  Results shown are obtained from: our theory (solid, dashed line); NSF equations (dot dashed cyan line); DSMC method (red square box)\cite{tantos2015conductive}.}
\end{figure}

We compare the results of two-temperature model with the Direct Simulation Monte Carlo (DSMC) method data \cite{tantos2015conductive}. Comparison between the analytical solution of
two-temperature model  and DSMC results are shown in Fig~\ref{density}, Fig~\ref{thetatr}, Fig~\ref{thetain}. Dimensionless
wall temperatures are at deviations of $\pm 0.0476$ from reference temperature at $350 K$. We
investigate two different reference Kn numbers, $0.071$ and $0.71$, which represent slip and
transition flow regimes, respectively. Also, excited internal
degrees of freedom is set to $2$, the same as the DSMC simulation. Fig~\ref{density}, Fig~\ref{thetatr}, Fig~\ref{thetain} clearly demonstrates the strong agreement between the DSMC and two-temperature model  results in slip flow regimes. Furthermore, in the transition regime, two-temperature model  exhibits even better agreement compared to the classical NSF theory. For $Kn=0.071$ and $Kn=0.71$, we determine the total heat flux ($\hat{q}_y$) to be approximately $0.025$ and $0.084$, respectively. As can be seen, the agreement between the curves of the two-temperature model solutions and the data from the DSMC reference solutions is evident, especially for smaller Knudsen numbers (Kn), indicating the accuracy and validity of two-temperature model  and boundary conditions.

\section{\label{Conclusions}Conclusions}
By integrating concepts from various approaches to Irreversible Thermodynamics, including LIT, RT, and RET, we have developed an enhanced set of constitutive relations for the stress tensor and heat flux within the framework of the two-temperature model.
The two-temperature model has been provided with boundary conditions that ensure thermodynamic consistency. These conditions include descriptions of velocity slip, temperature jump, and transpiration flow at the boundaries. \\
In this paper, a two-temperature model equation has been applied to the analysis of sound propagation, light scattering experiments, and heat transfer between two parallel plates in dilute polyatomic gases at room temperatures.\\
The comparison of our analytical solutions with acoustic and laser scattering experiments conducted in nitrogen, oxygen, carbon dioxide, and methane at room temperatures demonstrates the excellent performance of the two-temperature model equation proposed in this study. The model proves to be highly effective in accurately describing time-dependent phenomena in polyatomic gases, yielding satisfactory agreement with experimental observations. We also solved steady one-dimensional stationary heat conduction analytically with a set of 6-moment equations and compared the results with DSMC simulations. \\
In this paper, we have established the linear stability of the equations governing the two-temperature model for disturbances across all wavelengths or frequencies.\\
In subsequent research endeavors, there is potential for the development of second law preserving numerical techniques\cite{doi:10.1137/21M1417508} tailored for nonlinear flows. Additionally, alternative numerical approaches like the method of fundamental solutions\cite{rana_saini_chakraborty_lockerby_sprittles_2021,PhysRevE.108.015306} warrant investigation. Furthermore, the expansion of the two-temperature model to encompass more comprehensive hydrodynamic models, including CCR polyatomic \cite{rana_barve_2023} and extended moments models\cite{rahimi2016macroscopic}, is part of the forthcoming agenda.

 \begin{acknowledgments}
 AK gratefully acknowledges the financial support
from the Council of Scientific and Industrial Research
(CSIR) [File No.: 09/719(0120)/2020-EMR-I]. ASR acknowledges the financial support from the Science and
Engineering Research Board, India, through the Grants
No. SRG/2021/000790 and No. MTR/2021/000417.
 
% We wish to acknowledge the support of the author community in using
% REV\TeX{}, offering suggestions and encouragement, testing new versions,
% \dots.
 \end{acknowledgments}

\appendix

\section{Appendix}
%\begin{theorem}
The maximum entropy distribution function which maximizes the entropy \eqref{entropy equation} under the constraints \eqref%
{Moments density and velocity}, \eqref{moments translational energy} and \eqref{moments internal energy} takes the following form:%
\begin{equation*}
f_{\mathrm{6}}=\underset{\text{Maxwellian}}{\frac{\rho }{m}\underbrace{\frac{%
1}{\sqrt{2\pi \theta _{tr}}^{3}}e^{-\frac{C^{2}}{2\theta _{tr}}}}}\underset{%
\text{Gamma}}{\underbrace{\frac{1}{\Gamma \left( \frac{\delta }{2}\right) }%
\frac{1}{I}\left( \frac{I}{\theta ^{in}}\right) ^{\delta /2}e^{-\frac{I}{%
\theta ^{in}}}}}. 
\end{equation*}%
%Proof:%
\begin{proof}
The proof of the theorem employs the Lagrange multiplier method. To achieve this, we introduce a vector of multipliers $(\mathrm{\xi}_{0}, \mathrm{\xi}_{i}, \mathrm{\xi}_{1}, \mathrm{\xi}_{2})$ and define the corresponding functional:
\begin{eqnarray}
\mathsterling\left[ f\right]=&&-k_{b}\int f\ln \frac{f}{f_{0}}dCdI+\mathrm{\xi} 
_{0}\left\{ \rho -m\int fdCdI\right\}\nonumber \\&&
-\mathrm{\xi}_{i}m\int fC_{i}dCdI+\mathrm{\xi}_{1}\left\{ \frac{3}{2}\rho \theta
^{tr}-m\int f\frac{C^{2}}{2}dCdI\right\} \nonumber\\&&+\mathrm{\xi} _{2}\left\{ \frac{\delta }{
2}\rho \theta ^{in}-m\int fIdCdI\right\},
\end{eqnarray}
Taking the derivative of the functional $\mathsterling$ with respect to $f$ and setting it to zero
\begin{equation*}
\frac{\partial \mathsterling}{\partial f}=-k_{b}\left( 1+\ln \frac{f}{f_{0}}%
\right) -\mathrm{\xi} _{0}m-\mathrm{\xi}_{i}mC_{i}-m\mathrm{\xi} _{1}\frac{C^{2}}{2}%
-m\mathrm{\xi}_{2}I=0 ,
\end{equation*}%
Therefore the solution of the Euler–Lagrange equation $\frac{\partial \mathsterling}{\partial f}=0 $
 is given by:
\begin{equation}
f_6=f_{0}\exp \left(-1-\frac{\mathrm{\xi} _{0}}{R}-\frac{\mathrm{\xi}_{i}}{R}C_{i}-\frac{%
\mathrm{\xi}_{1}}{R}\frac{C^{2}}{2}-\frac{\mathrm{\xi}_{2}}{R}I\right),
\label{distribution}
\end{equation}
Put the value of $f_6$ in the constraint \eqref
{Moments density and velocity}
\begin{equation}
    \rho=m\int e^{-1-\frac{\mathrm{\xi}_0}{R}} I^{(\delta/2-1)}e^{-\frac{\mathrm{\xi}_1}{R}\frac{C^2}{2}} e^{-\frac{\mathrm{\xi}_2}{R}I}d  \mathbf{C}dI,
\end{equation}
\begin{equation}
    \rho=m e^{-(1+\frac{\mathrm{\xi}_0}{R})} A^{-(\delta/2)}\Gamma{(\delta/2)} \frac{2\sqrt{2}\pi^{\frac{3}{2}}}{B^{\frac{3}{2}}},
    \label{density l1}
\end{equation}
here $\delta>0$, B=$\frac{\mathrm{\xi}_1}{R}>0$ and A=$\frac{\mathrm{\xi}_2}{R}>0$.\\
Put the value of $f_6$ in the constraint \eqref{moments translational energy}
\begin{equation}
    \frac{3}{2}\rho \theta^{tr}=m\int e^{-1-\frac{\mathrm{\xi}_0}{R}} I^{(\delta/2-1)} e^{-\frac{\mathrm{\xi}_1}{R}\frac{C^2}{2}} e^{-\frac{\mathrm{\xi}_2}{R}I} \frac{ \mathbf{C}^2}{2}d  \mathbf{C}dI,
\end{equation}
\begin{equation}
    \frac{3}{2}\rho \theta^{tr}=m e^{-1-\frac{\mathrm{\xi}_0}{R}} \frac{3\sqrt{2}\pi^{\frac{3}{2}}}{B^{\frac{5}{2}}} A^{-(\delta/2)}\Gamma(\delta/2).
    \label{trans l2}
\end{equation}
Put the value of $f_6$ in the constraint \eqref{moments internal energy}
\begin{equation}
    \frac{\delta}{2}\rho \theta^{in} =m\int e^{-1-\frac{\mathrm{\xi}_0}{R}} I^{(\delta/2)} e^{-\frac{\mathrm{\xi}_1}{R}\frac{C^2}{2}} e^{-\frac{\mathrm{\xi}_2}{R}I}d  \mathbf{C}dI,
\end{equation}
\begin{equation}
    \frac{\delta}{2}\rho \theta^{in} =m e^{-1-\frac{\mathrm{\xi}_0}{R}} \frac{2\sqrt{2}\pi^{\frac{3}{2}}}{B^{\frac{3}{2}}} A^{-\left(1+\frac{\delta}{2}\right)}\Gamma\left(1+\frac{\delta}{2}\right).
    \label{inter l3}
\end{equation}
Solving equations \eqref{density l1}, \eqref{trans l2}, and \eqref{inter l3} we find the value of Lagrange multiplier as follows:
\begin{eqnarray}
   e^{-1-\frac{\mathrm{\xi}_0}{R}}&=&\frac{\rho}{m} \frac{1}{\sqrt{2\pi\theta^{tr}}^3}\left[\frac{1}{\theta^{in}}\right]^{\frac{\delta}{2}}\frac{1}{\Gamma(\delta/2)}, \\ \mathrm{\xi}_1&=&\frac{R}{\theta^{tr}}\text{, } \mathrm{\xi}_i=0\text{, and }
   \mathrm{\xi}_2=\frac{R}{\theta^{in}}.
\end{eqnarray}
Substitute these values in equation \eqref{distribution}, and we find the following maximum entropy distribution function
\begin{equation}
    f_6=\frac{\rho}{m} \frac{1}{\sqrt{2\pi\theta^{tr}}^3} e^{-\frac{C^2}{2\theta^{tr}}} \frac{1}{\Gamma(\frac{\delta}{2})}\frac{1}{I}\left[\frac{I}{\theta^{in}}\right]^{\frac{\delta}{2}} e^{-\frac{I}{\theta^{in}}}.
\end{equation}
\end{proof}
\bibliography{aipsamp}% Produces the bibliography via BibTeX.

\end{document}